\pdfoutput=1
\documentclass[onecolumn,a4paper,nopdfoutputerror]{quantumarticle}
\usepackage[utf8]{inputenc}
\usepackage[T1]{fontenc}
\usepackage{amsmath,amssymb}
\usepackage{booktabs}
\usepackage{graphicx}
\usepackage{bm}
\usepackage[numbers,sort&compress]{natbib}

\makeatletter
\AtEndPreamble{\RequirePackage{hyperref}}
\makeatother

\begin{document}

\title{Certifying bipartite entanglement on a superconducting processor from a corrected QAOA cost layer}

\author{Carlos-Miguel Lorenzo}
\orcid{0000-0003-2161-7301}
\email{carlos.lorenzo@cunef.edu}
\affiliation{CUNEF Universidad, Madrid, Spain}

\maketitle

\begin{abstract}
The Quantum Approximate Optimisation Algorithm is the canonical near-term heuristic, yet on real
superconducting hardware its behaviour is set by device physics, not by the algorithm. Hardware
demonstrations rarely prove that the intended coupling is realised, that the circuit prepares genuine
entanglement, or which physical parameter controls that entanglement; they report success against a random
floor. We report a pre-registered, provenance-controlled single-device study on the nine-qubit
superconducting processor Red at the Barcelona Supercomputing Center, using its five-qubit component.
Across six independent calibrations we certify bipartite entanglement of a corrected QAOA cost layer on the
good coupler at depth one by two-qubit tomographic negativity: day-level $0.077$,
bias-corrected-and-accelerated $95\%$ confidence interval $[0.065,0.091]$, excluding zero on all six days,
while the coupling-off twin stays separable and the degraded coupler is equivalent to zero. We then measure
the underlying device physics directly. In a pre-registered manipulation experiment on a fixed Bell
preparation --- not on the QAOA cost layer itself --- we deliberately vary a single per-job knob, the CZ
pulse amplitude, and certify the negativity at each setting. Across three couplers and two calibration
days, reducing the two-qubit drive causally reduces the certified negativity along a positive, well-fit
slope over the probed amplitude ladder ($R^2=0.83$--$0.96$), every rung-level (lack-of-fit-honest) $95\%$
interval excluding zero and the tests surviving Holm and Benjamini--Hochberg correction. This is a local
linearisation of a peaked response, not a global monotone law: the drive-to-phase map overshoots $\pi$ at
the nominal amplitude, so the certified negativity peaks at conditional phase $\lvert\zeta\rvert=\pi$ just
below the nominal rung in four of the six runs, rather than growing without bound. The drive-off $J=0$ twin is a separate off-state
control (zero amplitude, separable), not a point on the fitted line, and the mean fringe visibility is held
constant, so the effect is coherent rather than decoherence. The independently measured average gate fidelity
$F_\mathrm{avg}$ is \emph{non-monotone} in the drive because its Ramsey conditional phase wraps, so we
report it as a diagnostic and do not use it as the regressor. A per-circuit
statevector guardrail excludes the diagonal compilation fault that we disclose and repair (worst deviation
$8.4\times10^{-15}$). A randomized-measurement witness on Red is consistent with entanglement but
underpowered at the measured magnitude, a reported lesson in honest power. The certificate collapses by
depth three, and a pre-registered sixteen-instance ensemble at depth three shows no optimisation signal
(one-sided Wilcoxon $p=0.15$). We claim no advantage; the instances are classically trivial. The value is a
measured account of how a coupler's two-qubit drive controls the entanglement it can certify on one real
device.
\end{abstract}

\section{Introduction}\label{sec:intro}

The Quantum Approximate Optimisation Algorithm (QAOA)~\cite{farhi2014qaoa,hadfield2019qaoa} is the
canonical near-term heuristic for combinatorial optimisation and one of the most-studied testbeds of the
noisy intermediate-scale quantum (NISQ) era~\cite{preskill2018nisq,cerezo2021vqa,bharti2022nisq}. It
encodes a classical cost function as an Ising Hamiltonian~\cite{lucas2014ising,glover2018tutorial},
alternates a problem unitary with a mixer, and tunes a handful of angles to concentrate measurement
probability on low-energy bitstrings~\cite{zhou2020qaoa,brandao2018concentration}. A decade of theory has
mapped where the ansatz should and should not help: reachability deficits at shallow
depth~\cite{akshay2020reachability,akshay2021reachability}, concentration of the objective for typical
instances~\cite{brandao2018concentration}, and, on satisfiability and related classes, evidence both for
and against a practical edge~\cite{boulebnane2024satisfiability,shaydulin2024scaling}. In parallel, a body
of results has bounded what noise permits: barren plateaus and noise-induced flatness of the
landscape~\cite{wang2021noiseinduced}, and formal limitations of variational optimisation on noisy
hardware~\cite{stilckfranca2021limitations}. The gap this literature leaves is empirical and specific.
When a QAOA circuit runs on a particular superconducting chip, what does the device actually do to it?

Hardware QAOA studies typically answer a narrower question than they appear to. They report a success
probability, or an approximation ratio, on a single calibration, against a uniform or random-angle
floor~\cite{harrigan2021qaoa,weidenfeller2022scaling}. Three things are usually left unproven. First, that
the intended two-qubit coupling is realised on the device at all: a compiler can silently emit a
circuit whose two-qubit terms cancel, leaving a separable product circuit that still beats a random
floor because the single-qubit bias alone does. Second, that the circuit prepares genuine entanglement
rather than a classically correlated mixture: a raised success probability is not an entanglement
certificate. Third, that any reported number is stable across the daily recalibration cycle of a real
facility, rather than a single-shot artefact of one good day. Each of these is a place where a careful
referee can, and should, ask whether the headline is a measured property of the hardware or a property
written into the circuit by construction. This paper is built to answer that class of objection directly.

Our stance is certification, not benchmarking. We treat the QAOA cost layer as a state-preparation
circuit and ask a question with an unambiguous, device-independent answer: does the two-qubit reduced
state on a chosen edge carry entanglement, measured on the hardware and certified against a separable
baseline on the same hardware? Entanglement of a two-qubit mixed state is exactly detectable by the
positivity of the partial transpose (the Peres--Horodecki criterion), which is necessary and sufficient
in dimension $2\times2$~\cite{peres1996ppt,horodecki1996separability}. The negativity of the partial
transpose is a computable entanglement monotone, faithful for two
qubits~\cite{vidal2002negativity,horodecki2009entanglement}. We reconstruct the two-qubit reduced state
by full local tomography, project it to the nearest physical density matrix, and read the negativity with
a bootstrap confidence interval. This is a mature toolset~\cite{guhne2009entanglement}: our contribution
is to deploy it with pre-registration, per-circuit compilation guardrails and hierarchical statistics on
a live processor, and to say precisely, and only, what the data support.

This differs in kind from device benchmarking. Quantum volume and capability
measurement~\cite{cross2019quantumvolume,proctor2022measuring} summarise a processor's aggregate ability
to run random circuits of a given size; they are the right tools for comparing devices, but they do not
certify that a particular circuit prepared a particular entangled state. We ask the more local question,
and we make its answer falsifiable in the strong sense: a coupling-off twin, prepared and measured on the
same hardware in the same session, must return a separable state, and a compiled circuit that fails the
statevector guardrail is never run. The four load-bearing questions we set out to answer are therefore
sharp. Is the intended coupling realised on the device? Does the circuit prepare certified entanglement?
Which physical parameter gates that entanglement? And does the standard mitigation toolkit help or merely
relabel the signal? Each has a concrete, pre-registered test, and each is reported with its uncertainty
and its honest scope.

The certification of mixed-state entanglement without full tomography has matured in parallel, through
local randomized measurements and their partial-transpose moments~\cite{elben2020ppt,neven2021ptmoments}.
Estimating a few moments of $\rho^{T_B}$ from random single-qubit bases, using the classical-shadow
formalism~\cite{huang2020shadows,elben2022toolbox}, yields a witness $W=p_2^2-p_3$ whose positivity
certifies a non-positive-partial-transpose state without reconstructing $\rho$. We use this second route
as an independent cross-check. We also use it as a case study in honest reporting: at the small
entanglement magnitude realised on this device, the witness is underpowered at any feasible number of
measurement settings, and we say so, downgrading its status on the primary device to ``consistent with
entanglement'' rather than a certification, while showing that the same estimator is fully powered, and
clears its threshold, where the signal is larger.

The certificate is only half the story a device-physics paper should tell. The second half is which
physical parameter gates it. Superconducting processors are heterogeneous: two nominally equivalent
couplers on the same chip can differ by twenty percentage points of two-qubit gate
fidelity~\cite{krantz2019guide,proctor2022measuring,cross2019quantumvolume}. Our five-qubit component
carries, in one connected instance, a healthy coupler and a degraded one, so a single circuit hosts both
the entanglement cut and a within-chip good-versus-bad contrast. We test whether the certified
entanglement tracks the measured coupler fidelity, and we are careful about the logical status of that
test. A natural good-versus-bad edge contrast is an association, not a demonstrated cause; a genuine
causal statement needs the coupling to be manipulated, not merely observed. We therefore do both.
We report the within-chip association, and then we deliver the manipulation. In a separate, pre-registered
experiment we take a single physical knob --- the CZ pulse amplitude on a read-only master runcard --- and
deliberately vary the two-qubit drive of a fixed Bell preparation across a ladder of settings, certifying
the two-qubit negativity at each rung. Across three couplers and two calibration days this returns a
measured dose-response: reducing the CZ drive amplitude causally lowers the certified negativity along a
positive, well-fit slope over the probed amplitude window, with the CZ-off twin a separate off-state control
(not a point on the fitted line) and the fringe visibility held constant so the effect is coherent. The
slope is a local linearisation of a peaked, saturating response --- the certified negativity peaks at
conditional phase $\lvert\zeta\rvert=\pi$, which the nominal amplitude overshoots --- not a global monotone
law. The clean causal variable is the manipulated drive itself; the average
gate fidelity $F_\mathrm{avg}$, reconstructed independently from Ramsey fringes, is \emph{non-monotone} in
that drive because its conditional phase wraps, so we report it as a diagnostic and not as the regressor.
This is the paper's central device-physics result, and we state it with its scope: on a Bell preparation,
we measure how a coupler's two-qubit drive controls the entanglement it can certify.

Depth is the last axis. Each QAOA layer adds entangling gates a NISQ device cannot afford; deeper
circuits should decohere toward the maximally mixed state even as the noiseless ansatz improves. Error
mitigation is the standard response --- readout unfolding~\cite{nation2021m3,bravyi2021readout,maciejewski2020readout}
and zero-noise extrapolation~\cite{temme2017errormitigation,giurgicatiron2020zne,li2017efficient,larose2022mitiq,cai2023errormitigation,endo2018practical}
--- but mitigation can both fake and mask an entanglement signal, and recent work shows that zero-noise
extrapolation can produce artefactual ``improvements'' on hardware~\cite{koster2026zne}. We therefore
keep the raw, unmitigated measurement as the certificate throughout and report mitigation only alongside
it. Warm-starting, which front-loads classical information and reduces coherent evolution, is the one
intervention that changes the circuit rather than the post-processing~\cite{egger2021warmstart,tate2023warmstart,sack2021annealing};
we include it as a labelled arm.

This paper makes no claim of quantum advantage. The instances are five-variable tree Ising models,
solved instantly by belief propagation or dynamic programming; we report classical baselines and forbid
any speedup, scaling or solver claim by pre-registration. The contribution is a rigorous, reproducible
account of what one real device does to a corrected QAOA cost layer: the compilation proven correct by a
statevector guardrail, the entanglement then measured on the hardware by tomography and certified against
an on-device separable twin, the good-versus-degraded coupler separation stated as an interval-backed
within-chip contrast whose causal upgrade is pre-registered, and the within-instance depth collapse
together with a separate pre-registered honest negative on the depth-three ensemble. We open by
disclosing, not hiding, a compilation fault in our own prior five-qubit attempt: it is a strong motivation
for the guardrail that now protects every number here, and we treat it as a support control rather than a
co-headline.

Concretely, this paper contributes the following. (i) A hardware entanglement certificate for a
corrected QAOA cost layer: the two-qubit negativity on the good coupler at depth one, $0.077$ with a BCa
$95\%$ interval $[0.065,0.091]$ excluding zero on all six calibrations, with a coupling-off twin that
certifies zero on the same device. (ii) A per-circuit anti-regression statevector guardrail that excludes
the diagonal $\mathrm{CZ}\,R_Z\,\mathrm{CZ}$ compilation fault and holds every compiled cost layer to its
intended unitary (worst deviation $8.4\times10^{-15}$), turning a disclosed compilation fault into an
auditable compilation control; the coupling realised on the hardware is shown by the tomography and the
twin, not by this figure. (iii) A measured causal law, on a fixed Bell preparation and not on the QAOA
cost layer itself: across three superconducting couplers and two calibration days, deliberately reducing
the CZ pulse drive amplitude causally reduces the certified tomographic negativity along a positive,
well-fit slope over the probed amplitude ladder ($R^2=0.83$--$0.96$); with rung-level (lack-of-fit-honest)
standard errors every one of the six $95\%$ intervals excludes zero (one-sided $p<10^{-3}$ each, surviving
Holm and Benjamini--Hochberg correction across the three couplers), and the per-coupler slopes agree
($\approx 0.085$ of certified negativity per one percent of CZ amplitude, between-coupler heterogeneity
negligible). The slope is a \emph{local} linearisation of a peaked, saturating response, not a global
monotone law: the drive-to-phase map overshoots $\pi$ at the nominal rung, the certified negativity peaks at
conditional phase $\lvert\zeta\rvert=\pi$ (so the response is non-monotone across the nominal amplitude, the
negativity reading higher one step below nominal in four of the six runs), and extrapolating the slope to
zero amplitude is unphysical. The CZ-off $J=0$ twin is a separate off-state control (zero amplitude,
separable), not a point on the fitted line, and the fringe visibility is held constant so the effect is
coherent rather than decoherence. The independently measured average gate fidelity $F_\mathrm{avg}$ is non-monotone in the
drive (its Ramsey conditional phase wraps) and is reported only as a diagnostic, never as the regressor.
The observational within-chip good-versus-degraded contrast is reported separately as an association, and
the pre-registered paired test of that ordering is a secondary hypothesis that does not survive
multiplicity correction. (iv) A within-instance depth collapse on \texttt{b08}, where the certified
negativity present at depth one is separable by depth three; even the ideal negativity halves by depth
three while the hardware value reaches zero. (v) An independent randomized-measurement witness, reported
honestly as consistent with entanglement on the primary device, where it is underpowered at the measured
magnitude. We also report a pre-registered honest negative: the depth-three sixteen-instance ensemble shows
no optimisation signal. Throughout, the primary certificate and its controls carry bias-corrected
bootstrap intervals at a named inferential unit, every claim is tied to a sealed pre-registration, and the
whole pipeline is sealed and available for reproduction, so that every number regenerates from the raw shots. In one line, this is a
pre-registered, provenance-sealed single-device study that certifies entanglement of a corrected QAOA cost
layer and then measures, by direct manipulation, how a coupler's two-qubit drive controls that entanglement;
the certificate and the causal law are the two load-bearing results, and every other result --- the
depth collapse, the ensemble null and the randomized-measurement witness --- is subordinate and reported
with its exact scope.

\section{Device and problem}\label{sec:device}

\paragraph{Processor.} Red is a superconducting processor on the Barcelona Supercomputing Center (BSC)
quantum platform (MareNostrum Ona), accessed by SSH, SLURM and the Qililab control
library~\cite{qililab}; circuits are built and simulated with Qibo~\cite{efthymiou2021qibo}. The native
gate set is $\{R_Z, R_X, \mathrm{CZ}, M\}$ with virtual $Z$~\cite{mckay2017zgates}. We always read the
live connectivity from the runcard rather than from documentation. On the calibration current from
4~August 2026 the chip exposes nine calibrated qubits with seven CZ couplers,
$\{1\text{-}0,\,1\text{-}2,\,1\text{-}3,\,4\text{-}5,\,4\text{-}6,\,6\text{-}8,\,7\text{-}5\}$. The
nominal $3$-$4$ bridge is absent, so the device splits into two disconnected components,
$A=\{0,1,2,3\}$ and $B=\{4,5,6,7,8\}$. This study uses component $B$.

\paragraph{Why component B.} It carries in one connected five-qubit instance both a high-fidelity coupler,
the good edge $4$-$5$, and a degraded coupler, the edge $6$-$8$. A single circuit therefore hosts the
coupling, a certifiable entanglement cut on the good edge, and a within-chip good-versus-degraded coupler
contrast. The calibrated indices $\{4,5,6,7,8\}$ are the runcard-addressed labels; BSC maps them to
full-chip physical qubits $\{17,16,18,20,22\}$ in the calibration current as of 2026-08-17. This
calibrated-to-physical assignment can shift between calibrations, but BSC has confirmed (support ticket)
that the physical pair implementing the calibrated $4$-$5$ coupling has been unchanged since 4~August
2026, so the same physical edge underlies the six-day tomography (6--11~Aug) and the
randomized-measurement run (13~Aug). We nonetheless anchor every claim to the addressed and
guardrail-verified realised coupling, not to a physical-qubit label. The addressed coupler set on
component $B$ is byte-identical (SHA-256 over the CZ list) across all six daily runcard snapshots and the
randomized-measurement run.

\paragraph{Device health.} Device conditions come from BSC's daily autocalibration on the actual run
dates, mirrored locally with an integrity manifest. Across the six campaign days the calibrated $q_4/q_5$
mean $T_1\approx22/32\,\mu$s and $T_2\approx20/7\,\mu$s (the archive labels $T_2$ in ns, confirmed by
BSC). The good $4$-$5$ CZ-gate fidelity averaged $94.7\%$ with single-shot readout $\approx96\%$, while
the degraded $6$-$8$ edge averaged $74.7\%$ (edge $1$-$2$, $85.6\%$); on the randomized-measurement
calibration (13~Aug) the $4$-$5$/$6$-$8$ CZ fidelities were $94.9\%/74.0\%$. These are automated
calibration outputs and we use them as contextual device health, not as day-level regressors: every
day-level quantitative claim rests on the in-situ characterisation actually taken during the campaign
(per-qubit readout confusion in both directions and a Bell $Z$-correlator on each used edge), hashed and
bound to that day's job identifiers. BSC's per-qubit labelling shifts between calibrations, so an earlier
static descriptor of the degraded edge is superseded by these dated run-date values.

\paragraph{Instances.} The instances are random frustrated $2$-local Ising models on the native tree
edges of component $B$ (no SWAP routing), generated by a seeded, pre-registered generator whose master
seed (\texttt{20260804}) was committed and hashed before any QPU time. The generator applies structural
filters only --- unique ground state, spectral gap above a threshold, optimum neither trivial nor extreme,
and the load-bearing requirement that the $J=0$ twin optimum differs --- never solution-content filters.
Of $77$ candidates $61$ were rejected, an acceptance rate of $20.8\%$; exact ground truth, twin optimum
and noiseless entropies are sealed per instance. One pre-registered \emph{primary} instance, \texttt{b08},
receives the full factorial and the entanglement certificate. It was chosen by a criterion fixed before
execution --- the largest noiseless R\'enyi-2 entropy on the good cut at depth one --- not by outcome. Its
fields and couplings are $h=(0.984,0.193,0.344,0.947,-0.613)$ and, on the native edges,
$J_{4\text{-}5}=0.360$, $J_{4\text{-}6}=-0.148$, $J_{5\text{-}7}=0.171$, $J_{6\text{-}8}=0.895$
(instance record SHA-256 \texttt{ba4aea8f\ldots}); the frozen depth-one angles are $\gamma=1.909$,
$\beta=0.734$. The noiseless target is a near-maximally-entangled good cut: the ideal two-qubit negativity
on edge $4$-$5$ is $0.3196$ and on edge $6$-$8$ is $0.0922$. \emph{We label $0.3196$ ideal everywhere; it
is the $F\!\to\!1$ anchor of the fidelity-response regression, never presented as a hardware value.} A
pre-registered $N=16$ instance ensemble was measured at depth three for generalisation of the depth
ceiling. Ground truth is obtained by exact enumeration of the $32$ states; the instances are classically
trivial, and no advantage, hardness, scaling or solver claim is made.

\section{Compilation fault and the guardrail}\label{sec:bug}

We disclose, rather than bury, a compilation fault in a prior five-qubit attempt, because it is both the
reason all our earlier absolute numbers are withdrawn and the reason every number in this paper is
protected by an explicit control. The prior cost layer compiled each quadratic term as
$\mathrm{CZ}(i,j)\,R_Z(j,2\gamma J)\,\mathrm{CZ}(i,j)$. This product is all-diagonal and, since CZ and
$R_Z$ commute and $\mathrm{CZ}^2=I$, it equals $R_Z(j,2\gamma J)$ alone: a single-qubit phase with
\emph{no} ZZ coupling. The QAOA effectively ran as a separable, single-qubit circuit. Noiseless
reproduction of the as-run angles returns a product state whose success probability matches the prior
manuscript, whereas the correct coupling at those same angles gives a probability an estimated sixty-fold
smaller and a different mode. We therefore \emph{withdraw} every prior absolute number without
reservation.

The corrected cost layer realises $\exp(-i\gamma J\,Z_iZ_j)$ by Hadamard conjugation of the phase,
$\mathrm{CNOT}\,R_Z(2\gamma J)\,\mathrm{CNOT}$ with
$\mathrm{CNOT}=(I\otimes H)\,\mathrm{CZ}\,(I\otimes H)$, keeping two physical CZ per coupling so the
two-qubit gate budget is unchanged. Every submitted circuit then passes an anti-regression
\emph{statevector guardrail}: before any shot is spent, the compiled noiseless statevector is compared to
the exact $\exp(-i\gamma H_C)$ (times the mixer), and the job aborts on mismatch. This is a numerical
compilation control, evaluated on the noiseless statevector, that excludes the diagonal
$\mathrm{CZ}\,R_Z\,\mathrm{CZ}$ product and every other miscompilation before any hardware time is spent;
it is not, by itself, a measurement of the hardware. The bound is pre-registered at $10^{-9}$; the observed
worst deviation over the campaign was $8.4\times10^{-15}$ on the BSC toolchain (job 11484, 89 QAOA and
tomography circuits), roughly six orders of magnitude inside the bound, identical to machine precision
across all six calibration days, with $0$ aborts. The randomized-measurement circuits passed the same
pre-flight guardrail, with a sealed worst deviation of $8.33\times10^{-16}$.
This converts a silent failure mode into an auditable, per-circuit control, and it answers one half of the
first thing a referee should ask of a hardware entanglement claim: the compiled circuit encodes the
intended coupling, not the diagonal fault. The other half --- that the coupling is realised on the actual
hardware --- is answered not by this guardrail but by the on-device tomography together with the coupling-off
$J=0$ twin, which certifies zero on the same device and readout.

\section{Methods}\label{sec:methods}

\subsection{In-situ device characterisation and provenance}\label{sec:char}

Every day we measure the device conditions ourselves rather than trusting vendor-reported values. Each
run records the per-qubit readout confusion matrix in both directions (the probabilities
$p(1\,|\,0)$ and $p(0\,|\,1)$ from prepared computational basis states) and a Bell $Z$-correlator on each
used edge, both hashed and bound to that day's SLURM job identifiers. On day one, for example, the
readout confusion was $q_4\,(0.011,0.078)$, $q_5\,(0.060,0.049)$, $q_6\,(0.008,0.052)$,
$q_7\,(0.011,0.071)$, $q_8\,(0.007,0.032)$, and the Bell $Z$-correlator $P(\text{equal})$ was $0.894$ on
the good edge and $0.825$ on the degraded edge. Provenance is enforced per job: the live runcard is
copied to an immutable snapshot and hashed before connecting, the job runs against the copy, and each
result records the run identifier, date, host, runcard and snapshot SHA-256, the Qililab version, the
per-circuit guardrail deviation, the SLURM job identifier and the shot count. A single SHA-256 manifest
then binds each figure and number to its raw counts, analysis script, instance seed, runcard snapshot and
guardrail pass. Readout mitigation, where reported, uses positivity-preserving iterative Bayesian
unfolding rather than matrix inversion, so that a mitigated separable state cannot acquire a spurious
negative eigenvalue.

\subsection{Circuit and factorial}\label{sec:arms}

The state-preparation circuit is the corrected cost layer of Section~\ref{sec:bug} followed by the
transverse mixer $R_X(2\beta)$. Angles are frozen before execution from exact multi-start noiseless
optimisation (differential evolution~\cite{storn1997de}, three restarts, over the $2^5$ enumerated
states, with the correct CNOT-conjugated ZZ), never tuned on hardware; the global-optimum evidence is in
Appendix~\ref{app:angles}. The experiment is a factorial over coupling (ON versus the $J=0$ twin, matched
in single-qubit depth), embedding (good edge $4$-$5$ versus the degraded $6$-$8$ edge), start (cold versus
warm), mitigation (none, readout, readout-plus-ZNE), and depth $p\in\{1,2,3\}$, repeated on at least six
calibration days. The warm-start arm uses the regularised Egger--Mare\v{c}ek--Woerner
scheme~\cite{egger2021warmstart,tate2023warmstart} at $\varepsilon=0.25$ with the warm-start mixer, added
as a dated, pre-data amendment before any warm-start data were taken. The scheme seeds each qubit from a
regularised continuous relaxation of the Ising problem and replaces the transverse mixer with a
relaxation-aligned one; we constrain the seed so that its rounded relaxation differs from the true optimum
on at least one load-bearing bit for every instance, so warm-starting biases the circuit toward a
partially wrong starting point that the coupling must still correct, and never hands QAOA the answer. This
matters because a warm-start optimum-mass number that merely reflected a near-optimal initialisation would
be uninformative; our construction removes that confound by design, which is why the standard, cold arm
remains the clean contrast for the optimisation tests. Submission order is randomised within a
day and blocked on day. In this reduced pilot the mitigation ladder ran to readout unfolding; the
ZNE-folding rung was not executed and is labelled deferred.

\subsection{Entanglement protocol}\label{sec:ent}

\paragraph{Primary certificate: tomographic negativity.} The primary certificate is the
negativity~\cite{vidal2002negativity} of the two-qubit reduced state on the certified cut, obtained from
full two-qubit tomography (nine Pauli-pair settings, $4096$ shots per setting, $36{,}864$ shots per arm
per day). We reconstruct $\rho$ by linear inversion and then project to the nearest physical (positive
semidefinite, unit-trace) density matrix before evaluating the negativity; the negativity is positive
exactly when the partial transpose is non-positive, which for two qubits is necessary and sufficient for
entanglement~\cite{peres1996ppt,horodecki1996separability}. The raw, unmitigated value is the
certificate. We report readout-mitigated values (positivity-preserving iterative Bayesian unfolding,
IBU~\cite{nation2021m3}) only alongside the raw value, because readout mitigation can both fabricate and
suppress an entanglement signal. The negative control is the coupling-off $J=0$ twin: a genuine
certificate must certify on the coupled arm and \emph{not} certify on the twin, on the same device and
readout.

We follow the estimator validated in the A5 recompute. On this data set the raw linear-inversion $\rho$
is already physical --- its minimum eigenvalue is positive on all six days on the good edge (of order
$+0.05$) --- so projecting to the nearest physical state leaves every point estimate \emph{unchanged}
(grand-mean shift $0.0000$); the projection only guarantees a physical matrix and removes a
false-certification risk at lower fidelity. What the recompute changes is the interval method: we replace
the percentile bootstrap, which miscovers a boundary statistic constrained to be non-negative, with a
bias-corrected-and-accelerated (BCa) bootstrap~\cite{efron1979bootstrap}, taking the bias correction from
the bootstrap distribution and the acceleration from a delete-one-shot-block jackknife (all nine Pauli
settings are structurally required, so a leave-one-setting-out jackknife is invalid). A self-check
confirms the estimator returns $0.5$ on a Bell state and $0.0$ on a product state and on the maximally
mixed state.

\paragraph{Second monotone, same data.} From the same reconstructed $\rho$ we also report the Wootters
concurrence~\cite{wootters1998concurrence} and the logarithmic negativity as robustness cross-checks, and
the CHSH-operator maximum of the reconstructed state as context. We state plainly that the concurrence is
\emph{not} an independent method: it is a second monotone computed from the same tomography, and it is
reported to show that the certificate does not depend on the choice of monotone. The CHSH-operator maximum
is a property of the reconstructed state under the Horodecki
criterion~\cite{horodecki1996separability,clauser1969chsh}, not a device-independent Bell test, and we
label it as such wherever it appears. These three secondary monotones carry day-level intervals rather than
the BCa bootstrap; the BCa interval is applied to the primary certificate and its two controls, where the
boundary-statistic miscoverage it corrects is load-bearing, and is not claimed universally.

\paragraph{Independent route: randomized-measurement witness.} As an independent cross-check we compute
the $p_3$-PPT mixed-state moment witness from local randomized
measurements~\cite{elben2020ppt,neven2021ptmoments} on the good-edge $4$-$5$ cut, on a subsequent Red
calibration (2026-08-13, jobs 11608--11611, runcard SHA \texttt{567afc2a}). We use single-qubit Clifford
classical shadows~\cite{huang2020shadows,elben2022toolbox}: the single-qubit Clifford group is a unitary
$3$-design, so the shadow-inversion channel $3(\cdot)-\mathrm{Tr}(\cdot)\,I$ is exact, and the
partial-transpose moments $p_2=\mathrm{Tr}((\rho^{T_B})^2)$ and $p_3=\mathrm{Tr}((\rho^{T_B})^3)$ are
estimated by distinct-index U-statistics over the random settings, unbiased by the independence of the
settings. The witness $W=p_2^2-p_3>0$ certifies a non-positive-partial-transpose state. We estimate
$p_2^2$ with its own distinct-quadruple U-statistic, because the plug-in $\hat p_2^{\,2}$ is biased upward
and would inflate $W$ for a separable state. The budget is $N_U=800$ settings by $N_M=400$ shots, i.e.\
$320{,}000$ shots.

\paragraph{Why the randomized-measurement witness is a cross-check, not the primary certificate.} The
decision rule is pre-registered and false-positive-controlled: a threshold at the $95$th percentile of
$\hat W$ under a mixed PPT-boundary null (Werner state, $p=1/3$, $W=0$), certifying only if the hardware
$W$ exceeds it. Against that boundary null the hardware witness is many standard deviations clear (see
Results). But the honest null for a high-purity separable state is not the Werner boundary; it is the
near-pure-separable family, whose finite-sample moment variance is large. A dedicated power study shows
that, at the entanglement magnitude realised on Red (Werner-equivalent $W\approx0.016$), the $95$th
percentile of $\hat W$ under that family exceeds the signal at every feasible number of settings: a
threshold below the measured signal would need $N_U\approx12{,}672$, against the $800$ actually run, and
the detection power at feasible budgets is essentially zero. We therefore report the randomized-measurement
witness on Red as \emph{consistent with entanglement}, not as a certification, and name the PSD+BCa
tomographic negativity as the single primary certificate. The same power study confirms the witness is
fully powered at the ideal magnitude ($W\approx0.245$, power $\approx1$), so it is fit for
large-entanglement cells; we exploit exactly that regime in the cross-architecture corroboration
(Section~\ref{sec:limits}). Line-3 mechanism observables (connected correlators, per-edge mutual
information) and the global-purity control were not processed in this pilot and are deferred.

\subsection{Causal coupler-fidelity experiment}\label{sec:causalmethods}

The within-chip good-versus-degraded contrast (Section~\ref{sec:r2}) is observational. To convert it into
a causal statement we designed a separate confirmatory manipulation experiment, in which the coupler
drive is set by us rather than read off the device. The design was pre-registered as a
kill-safe, sealed protocol, and each day's plan was sealed with an independent external
timestamp \emph{before} that day's data existed (Section~\ref{sec:char}; tokens in
Appendix~\ref{app:B}). The pre-registered decision rule is one-sided: the run is a GO if the fitted slope
is positive \emph{and} its $95\%$ confidence interval excludes zero, and a pre-committed null otherwise
(we report the interval on the corrected axis with rung-level lack-of-fit-honest errors,
Section~\ref{sec:deviations}).

\paragraph{Intervention.} The target state is a fixed maximally-entangled Bell preparation
$\lvert\Phi^+\rangle$ (Hadamards and one CZ on the chosen edge), identical across the ladder. The single
manipulated variable is the CZ pulse amplitude, applied through a surgical one-line scale on a per-job
\emph{copy} of the runcard; the master runcard is read-only and its SHA-256 is re-verified identical
before and after every rung, so nothing else on the chip changes. The injected error is therefore purely a
\emph{coherent} conditional-phase detuning of the flux pulse, not added stochastic noise.

\paragraph{Response axis: the manipulated drive, not the gate fidelity.} The causal regressor ($x$-axis)
is the single manipulated variable itself --- the CZ pulse amplitude factor, the dimensionless fraction of
the nominal drive that we set on the runcard --- equivalently the measured conditional phase
$\lvert\zeta\rvert$ (with $\zeta=\phi_{ab}-\phi_a-\phi_b$ reconstructed by three Ramsey fringes on
independent equatorial input states) as the physical entangling parameter. The per-rung $\zeta$ is measured
modulo $2\pi$; we report the physically \emph{unwrapped} $\lvert\zeta\rvert$ (phase unwrapped along the
amplitude ladder), which is a single-valued monotone function of the drive up to $\pi$. The $y$-axis is the same
PSD-constrained tomographic negativity used for the certificate (nine Pauli-pair settings, projection to
the nearest physical state), reusing the A5 estimator without modification. The amplitude is set on the
runcard and the negativity is read from tomography, so the $x$ and $y$ measurements are independent. We
deliberately do \emph{not} regress on the coherent average gate fidelity $F_\mathrm{avg}=(4F_e+1)/5$: as
the amplitude is swept, the conditional phase $\zeta$ passes through $\pi$ and \emph{wraps}, so
$F_\mathrm{avg}$ --- a function of $\cos\zeta$ through $F_e$ --- is \emph{non-monotone} in the drive and
does not track the certified negativity (which rises with $\lvert\zeta\rvert$ up to its entangling peak at
$\lvert\zeta\rvert=\pi$). $F_\mathrm{avg}$ is therefore reported as a diagnostic only
(Appendix~\ref{app:B}), never as the regressor. The mean fringe visibility is reported separately as a
decoherence diagnostic, precisely so that a coherent effect can be distinguished from a loss of visibility.

\paragraph{Ladder, anchor and rung-level inference.} A short fine sweep first locates a monotonic segment
of $\zeta$ near the nominal amplitude; eight rungs are then placed evenly in $\zeta$ across that segment,
anchored at the nominal amplitude, and submitted in a single batch so they share one calibration nominal.
A CZ-off rung (amplitude zero, the $J=0$ twin) is included as a pre-committed \emph{off-state control}
(zero drive, zero certified negativity); it is a qualitative anchor, not one of the eight fitted rungs, and
is not a point on the regression line. The response is a weighted least-squares regression of the
certified negativity on the manipulated CZ amplitude factor (equivalently on the unwrapped
$\lvert\zeta\rvert$) over the eight rungs, with weights equal to the inverse bootstrap variance of the
per-rung negativity. Because the eight rungs span only a narrow window near the nominal amplitude, over
which the drive-to-phase map is monotone up to $\pi$, the fitted slope is a \emph{local} linearisation of a
peaked, saturating response and must not be extrapolated to zero amplitude (its intercept is large and
negative, predicting an unphysical negativity there). Crucially, the slope
uncertainty is computed at the \emph{rung} level, not from the shot bootstrap: the eight rungs scatter
about the fitted line by more than their shot noise (reduced $\chi^2>1$), reflecting the residual curvature
of a locally linear fit to a saturating response, so we inflate the model (shot-level) slope standard error
by $\sqrt{\smash[b]{\chi^2_\nu}}$ and take a Student-$t$ interval on $\nu=n-2=6$ degrees of freedom. A
shot-level bootstrap that treats the rungs as exact would understate the uncertainty; we report $R^2$ and
the reduced $\chi^2$ (lack of fit) for every fit. The monotonicity check on the fine $\zeta$ segment is a
rung-\emph{placement} diagnostic, \emph{not} the pre-registered decision rule: the kill-gate is that the
fitted slope is positive \emph{and} its interval excludes zero. On one run (edge $5$-$7$, session~2) the
pre-scan flagged a non-monotone $\zeta$ at the smallest amplitude
(\texttt{monotonic\_segment.monotonic\_all}${}=$\texttt{false}), so its ladder stops one amplitude step
earlier (smallest factor $0.976$ rather than $0.972$); the eight placed rungs are themselves monotone and
the run meets the kill-gate on the corrected axis (Appendix~\ref{app:B}). Every run carries the same
per-job guardrail, provenance and self-checks as the main campaign (tomography returns $0.5$ on a Bell
state and $0.0$ on a product state; the fringe fitter recovers injected phases).

\paragraph{Replication.} The experiment was run on three physical couplers of component $B$ ---
$4$-$5$, $5$-$7$ and $6$-$8$ --- on two independent calibration days with distinct master runcards
(session~1 master SHA-256 \texttt{e8db3ca5\ldots}, session~2 \texttt{d597a096\ldots}), giving six
independent runs. Per-edge slopes pool the two sessions by inverse-variance weighting; the edge-level
random-effects summary and the multiplicity control across the three couplers are defined in
Section~\ref{sec:stats}.

\subsection{Statistics and inference}\label{sec:stats}

\paragraph{Inferential units.} The unit hierarchy is shot $<$ run $<$ day $<$ instance; for the
randomized-measurement estimator the unit is the random-unitary setting. Shots are never treated as
independent replicates for cross-run, cross-day or cross-instance claims. Every primary estimate carries a
confidence interval at its named unit, with an effect size on every paired test. The pre-registered
analysis plan is a beta-binomial generalised linear mixed model plus hierarchical BCa bootstrap with a
four-way variance decomposition; in this pilot the certificate statistics (the negativity, the coupler
contrast and the day-level stability) use the hierarchical day-level bootstrap over the six calibration
days, and the ensemble optimum-mass test uses an instance-level paired one-sided Wilcoxon across the
$N=16$ instances. The mixed model, the full variance decomposition and the out-of-sample noise model are
pre-registered for the depth-one-and-two ensemble run and reported there, not here.

\paragraph{Nulls and controls.} Two controls anchor the certificate. The coupling-off $J=0$ twin is the
named specificity control; the degraded $6$-$8$ edge is the second control. Both are assessed by
two-one-sided-tests (TOST) equivalence to zero at a pre-registered margin $\delta=0.02$, not merely by a
non-significant point estimate. For optimum-mass, the pre-registered nulls are a spectrum-preserving
scrambled-encoding null and a device-marginal Bernoulli null, with the uniform floor ($1/32=0.03125$) and
a random-angle ensemble as floors. In this pilot the ensemble optimum-mass was tested against the
device-marginal Bernoulli null; the scrambled-encoding hardware null was not executed and that hypothesis
is therefore partially tested.

\paragraph{Estimands and effect sizes.} The certificate estimand is the day-level grand-mean negativity,
with the day as the top random effect; the ensemble estimand is the per-instance optimum mass
$P_\mathrm{opt}$, the probability that a single measurement returns the unique optimal bitstring, and the
approximation ratio $r=(\langle E\rangle - E_\mathrm{u})/(E^\star - E_\mathrm{u})$ referenced to the
uniform mean $E_\mathrm{u}$ and the ground energy $E^\star$. Paired tests carry a rank-biserial
correlation as the effect size; the pre-registered analysis plan additionally specifies a beta-binomial
generalised linear mixed model with a four-way variance decomposition (shot, run, day, instance) for the
full ensemble run, which is reported there rather than in this reduced pilot. The intraclass correlation
at the day level, quoted for the certificate as a stability descriptor, is the fraction of total
negativity variance attributable to the between-day component.

\paragraph{Randomized-measurement estimator.} For the witness, the partial-transpose moments are
distinct-index U-statistics over the random single-qubit-Clifford settings: $p_2$ from ordered pairs of
distinct settings, $p_3$ from ordered triples, and $p_2^2$ from ordered quadruples, so that each is an
unbiased estimator of its target under the independence of the settings. The witness is
$W=\widehat{p_2^2}-\hat p_3$, and using the quadruple U-statistic for $p_2^2$ rather than the square of
the $p_2$ estimate is what preserves the one-sided guarantee in finite samples: the plug-in
$\hat p_2^{\,2}$ is biased upward and would inflate $W$ for a separable state. The decision threshold is
the $95$th percentile of the bootstrap $\hat W$ under a Werner PPT-boundary null; the interval is a BCa
bootstrap over the $N_U$ settings. The estimator, its brute-force cross-checks and its operating
characteristic were validated in simulation before any QPU time.

\paragraph{Causal dose-response inference.} For the manipulation experiment
(Section~\ref{sec:causalmethods}) the per-run estimand is the slope of the certified negativity on the
\emph{manipulated} CZ pulse amplitude factor (equivalently on $\lvert\zeta\rvert$), fitted per run by
weighted least squares with the rung-level lack-of-fit-honest interval defined above (shot-level slope
standard error inflated by $\sqrt{\smash[b]{\chi^2_\nu}}$, Student-$t$ on $\nu=6$), and reported with its
$R^2$ and reduced $\chi^2$. We then pool at two levels. Each coupler's slope pools its two sessions by
inverse-variance weighting of the two rung-honest session slopes, giving a per-coupler slope and a
normal-theory interval. Across the three couplers we report a random-effects (DerSimonian--Laird) pooled
slope with the edge as the random unit, quoting both the standard normal interval and, because only three
couplers enter, the conservative small-sample Knapp--Hartung--Sidik--Jonkman interval; we report the
between-coupler heterogeneity ($Q$, $I^2$, $\tau^2$) so that the model dependence of the pooled magnitude
is explicit. Multiplicity across the three couplers is controlled by Holm and Benjamini--Hochberg
correction of the one-sided per-coupler slope tests. The causal claim is designed to rest on the per-coupler
evidence and its unanimity; on the manipulated axis the pooled magnitude is also well behaved (the
per-coupler slopes agree), and we report it with its heterogeneity so the reader can judge. The re-analysis
script refits every per-run slope from the raw per-rung amplitude and negativity arrays as a self-check.

\paragraph{Finite statistics.} The tomography is expensive and single-instance: $36{,}864$ shots per arm
per day, $221{,}184$ shots over six days for the certificate cell alone, plus matched budgets on the twin,
the degraded edge, the warm arm and the depth-three arm; the randomized-measurement cross-check adds a
$320{,}000$-shot budget. We report the certificate, its controls and the causal slopes as BCa intervals at
the day, setting or rung level, and the secondary tomographic monotones as day-level intervals, never as a
shot-count standard error. The day-level stopping rule was pre-registered: add days until the
day-level grand-mean CI half-width falls to $0.02$ or below; the six-day window satisfies it (half-width
$0.017$). The randomized-measurement estimator's unbiasedness and false-positive behaviour were validated
in simulation (null mean $W$ within $10^{-6}$ of zero, false-positive rate of the CI-excludes-zero rule
$1.5\%$) and by a closed-loop self-check on the exact compiled circuit, all before any QPU time.

\subsection{Deviations from pre-registration}\label{sec:deviations}

We depart from the sealed plan in five ways, each disclosed here rather than left for a reader to
reconstruct. \emph{(a) The primary endpoint is inverted.} The sealed \texttt{REGISTRATION.md} names, as
primary hypothesis PP1, the $p_3$-PPT randomized-measurement moment witness as the entanglement
certificate, with the two-qubit tomographic negativity as its cross-check. We invert that order and make
the negativity the single primary certificate, because the randomized-measurement witness is underpowered
at the entanglement magnitude realised on Red: at $W\approx0.016$ the honest near-pure-separable null is
not controllable below the signal until $N_U\approx12{,}672$ settings, against the $N_U=800$ actually run,
so no threshold there both controls the false-positive rate and retains power. The witness is retained in
its powered regime and reported as a corroboration, never as the on-Red certificate.
\emph{(b) The interval estimator changed.} The pre-registered interval was a percentile bootstrap; we
report a bias-corrected-and-accelerated (BCa) bootstrap instead, because the negativity is a boundary
statistic constrained to be non-negative and the percentile interval miscovers it. On this data set the
change moves the bounds by about $2\times10^{-4}$ and does not alter any certification decision.
\emph{(c) The executed shot budget is below the pre-registered target.} The plan specified
$\geq 8192$ shots per tomography setting for a Wilson half-width near $0.006$; this reduced pilot executed
$4096$ shots per setting. The day-level stopping rule (add days until the grand-mean CI half-width
$\leq0.02$) was nonetheless met at six days (half-width $0.017$), and every uncertainty is reported at the
day or setting level rather than as a shot-count standard error.
\emph{(d) The multiplicity family is stated explicitly.} The pre-registration controls the secondary
hypotheses H3--H9 (seven hypotheses) by a Holm/BH procedure within the family. Four were evaluable in this
pilot. The strongest nominal secondary is H4 (good edge $>$ degraded edge), a paired one-sided Wilcoxon
with raw $p=0.0156$; the Holm step-down threshold for the first rejection is $0.05/7=0.0071$ over the full
family, or $0.05/4=0.0125$ over the four tested, and $0.0156$ exceeds both, so H4 does not survive
correction. At $n=6$ the one-sided Wilcoxon is in any case pinned at its floor $1/2^{6}=0.0156$, which
carries one bit of sign information, not a magnitude. The remaining evaluable secondaries (warm versus
cold, the depth collapse, the day-level variance component) are reported descriptively with intervals and
reach no smaller nominal $p$. We therefore do not present any secondary as a corrected-significant result.
\emph{(e) The causal regressor is the manipulated drive, not the measured gate fidelity.} The sealed
manipulation protocol regressed the certified negativity on the measured coherent CZ fidelity
$F_\mathrm{avg}$, and each of the six sealed records carries a \texttt{GO} verdict evaluated on \emph{that}
$F_\mathrm{avg}$ fit with a shot-level bootstrap. We refit on the manipulated CZ amplitude itself
(equivalently on the unwrapped conditional phase $\lvert\zeta\rvert$), because $F_\mathrm{avg}$ is
non-monotone in the drive --- its Ramsey conditional phase wraps through $\pi$ --- so a slope fitted to it
is not interpretable and its shot-level interval understates the rung-to-rung lack of fit
(negativity-vs-$F_\mathrm{avg}$: weighted $R^2\le0.18$, reduced $\chi^2\approx100$--$137$, and no
per-session slope excludes zero with rung-honest errors). This is a \emph{principled post-hoc} correction on
a data-independent physics ground: the non-monotonicity is a property of the drive-to-phase map, not of the
measured outcome, and would hold had the run not been taken --- the same logic as the endpoint elevation of
item~(a) (deposit note \texttt{AMENDMENTS.md}, section~3). On the manipulated axis the response is a
positive, well-fit slope over the probed ladder and every run meets the pre-registered kill-gate (positive
slope, interval excluding zero) with rung-honest errors. We therefore do \emph{not} read the sealed
$F_\mathrm{avg}$-axis \texttt{GO} as corroborating the corrected-axis fit: the pre-registered gate is met on
the sealed $F_\mathrm{avg}$ fit and, \emph{independently}, on the corrected amplitude and $\lvert\zeta\rvert$
axes. $F_\mathrm{avg}$ is retained as a reported diagnostic (Appendix~\ref{app:B}).

\section{Results}\label{sec:results}

We name the inferential unit with every number. The certificate is the raw negativity; readout-mitigated
values appear in parentheses as context. Day-level statistics are a hierarchical bootstrap over the six
calibration days; ensemble statistics are instance-level across $N=16$.

\begin{figure}[t]
\centering
\includegraphics[width=\linewidth]{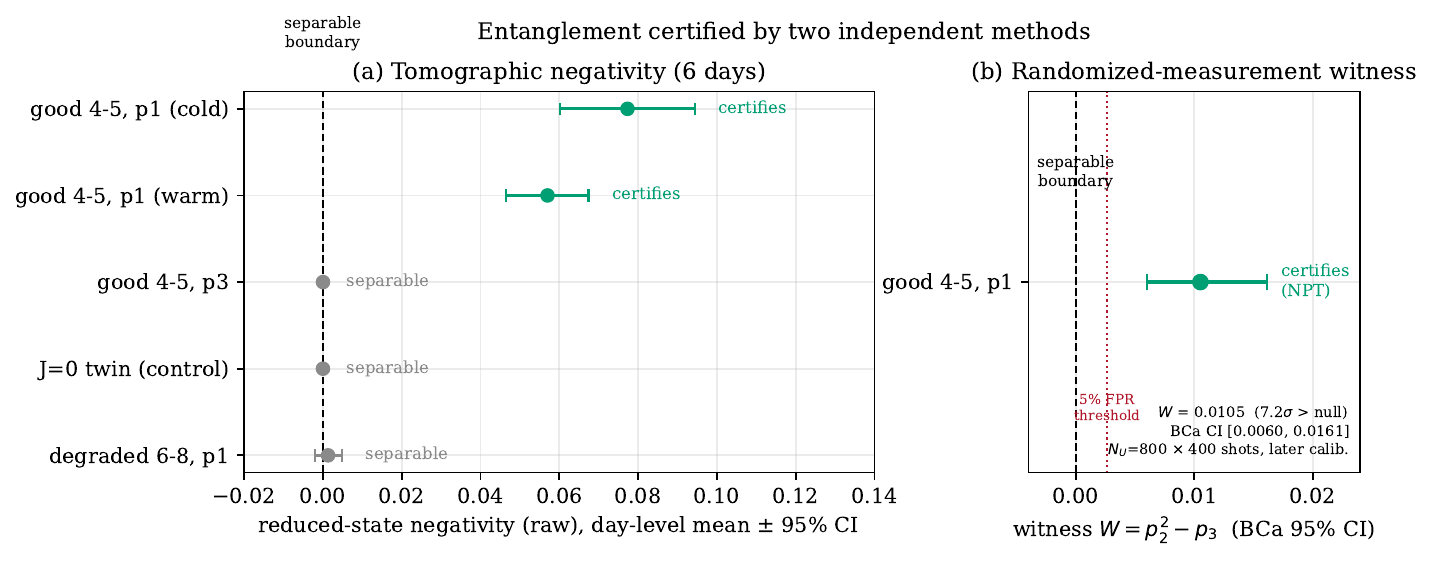}
\caption{\textbf{Entanglement is certified on the good coupler at depth one and nowhere else.} (a)
Per-arm two-qubit reduced-state negativity from tomography, day-level mean with $95\%$ confidence
interval over six calibrations: the good edge $4$-$5$ at $p=1$ certifies (cold and warm), while the
$p=3$ arm, the coupling-off $J=0$ twin and the degraded $6$-$8$ edge are separable. (b) The independent
randomized-measurement $p_3$-PPT witness $W=p_2^2-p_3$ on the good edge on a later calibration:
$W=0.0105$ with BCa $95\%$ CI, clear of the Werner PPT-boundary null but --- at this magnitude ---
underpowered against a near-pure-separable null, so reported as consistent with entanglement.}
\label{fig:certificate}
\end{figure}

\subsection{Entanglement is certified on the good coupler at depth one}\label{sec:r1}

On the good edge $4$-$5$ at depth one, on the primary instance \texttt{b08}, the coupling is live and the
two-qubit reduced state is entangled. The certificate is the PSD-constrained tomographic negativity: a
day-level grand mean of $\mathbf{0.077}$ with a BCa $95\%$ confidence interval $\mathbf{[0.065, 0.091]}$
that excludes zero, over six independent daily calibrations (between-day standard deviation $0.016$;
per-day values $0.104,\,0.072,\,0.084,\,0.078,\,0.055,\,0.070$; each day's own interval excludes zero, so
the certificate holds on $\mathbf{6/6}$ days). The more conservative day-level $t$-interval is
$[0.060,0.094]$, also excluding zero. Because the raw reconstruction is already physical on every day, the
projection to the nearest physical state shifts the point estimate by $0.0000$: the certificate is a
property of the measured state, not of the reconstruction choice. Readout mitigation raises the grand mean
to $0.152$ (BCa $95\%$ CI $[0.138,0.170]$), reported only as context; the raw value is the claim.

A second monotone from the same tomography agrees. The Wootters concurrence has day-level grand mean
$\mathbf{0.165}$ (day-level CI $[0.131,0.200]$), and the logarithmic negativity $0.207$ (day-level CI
$[0.164,0.250]$), both excluding zero on the good edge. The CHSH-operator maximum of the reconstructed
state is $1.37$ (day-level CI $[1.31,1.44]$), below the value $2$: the state is entangled but not, at this
fidelity, capable of a CHSH violation --- reported honestly as a state bound, not a device-independent
Bell test. These three secondary monotones are robustness cross-checks and we report them with day-level
intervals; the BCa treatment is reserved for the primary certificate and its controls (the negativity, the
$J=0$ twin and the degraded edge), for which it is load-bearing.

The controls do exactly what a certificate requires. The coupling-off $J=0$ twin returns negativity
$\mathbf{0.00}$ on both monotones and is TOST-equivalent to zero at margin $0.02$ (upper $95\%$ bound
$0.0001$). The degraded $6$-$8$ edge does not certify: negativity $\mathbf{0.0013}$ with a non-negativity-respecting
BCa $95\%$ CI $[0.000,0.007]$ that includes zero (concurrence $0.005$, CI including zero), also
TOST-equivalent to zero (upper $95\%$ bound $0.0046$). We report the BCa interval rather than the
day-level $t$-interval $[-0.0021,0.0048]$ for the controls, because negativity is constrained to be
non-negative and a negative lower bound is not a meaningful uncertainty for it. The certificate is
single-instance (\texttt{b08}) and rests on the raw, unmitigated measurement against an on-device
separable control (Fig.~\ref{fig:certificate}). Table~\ref{tab:perday}
gives the day-by-day values behind these summaries.

\begin{table}[t]
\centering
\caption{\textbf{Per-day two-qubit negativity behind the certificate.} Good edge $4$-$5$ at $p=1$ (raw
and readout-mitigated), the coupling-off $J=0$ twin, the degraded $6$-$8$ edge, and the warm-start good
edge, over six independent daily calibrations (raw is the certificate). Each good-edge day excludes zero;
the twin and the degraded edge do not. The final rows give the day-level grand mean, between-day standard
deviation and BCa $95\%$ confidence interval.}
\label{tab:perday}
\begin{tabular}{lccccc}
\toprule
Calibration & good $4$-$5$ raw & good $4$-$5$ (IBU) & $J{=}0$ twin & bad $6$-$8$ & warm good $4$-$5$ \\
\midrule
Day 1 & $0.104$ & $0.188$ & $0.000$ & $0.000$ & $0.065$ \\
Day 2 & $0.072$ & $0.156$ & $0.000$ & $0.000$ & $0.044$ \\
Day 3 & $0.084$ & $0.149$ & $0.000$ & $0.000$ & $0.068$ \\
Day 4 & $0.078$ & $0.149$ & $0.000$ & $0.008$ & $0.063$ \\
Day 5 & $0.055$ & $0.128$ & $0.000$ & $0.000$ & $0.054$ \\
Day 6 & $0.070$ & $0.143$ & $0.000$ & $0.000$ & $0.048$ \\
\midrule
Grand mean & $\mathbf{0.077}$ & $0.152$ & $0.000$ & $0.001$ & $0.057$ \\
Between-day SD & $0.016$ & $0.020$ & $0.000$ & $0.003$ & $0.010$ \\
BCa $95\%$ CI & $[0.065,0.091]$ & $[0.138,0.170]$ & TOST${=}0$ & $[0.000,0.007]$ & $[0.048,0.066]$ \\
\bottomrule
\end{tabular}
\end{table}

\subsection{A within-chip separation between the good and degraded couplers}\label{sec:r2}

The good edge certifies and the degraded edge does not, and the separation is consistent day by day. The
certified content is the pair of per-edge day-level intervals: the good edge $4$-$5$ excludes zero on
$\mathbf{6/6}$ days with grand-mean negativity $0.077$ (BCa CI $[0.065,0.091]$), while the degraded edge
$6$-$8$ is TOST-equivalent to zero (grand mean $0.0013$, BCa CI $[0.000,0.007]$). The good edge is ahead on
$\mathbf{6/6}$ days, and the device-health Bell $Z$-correlator is correspondingly higher on the good edge
($\approx0.90$ versus $\approx0.84$, six-day means). The pre-registered paired test of this ordering (H4,
good edge $>$ degraded edge) is a \emph{secondary} hypothesis, and we report it as such: the one-sided
Wilcoxon signed-rank test gives raw $p=0.0156$, which does \emph{not} survive Holm/BH multiplicity
correction across the pre-registered secondary family (first-rejection threshold $0.05/7=0.0071$ over the
seven secondaries, $0.05/4=0.0125$ over the four evaluable). At $n=6$ the one-sided Wilcoxon is in any case
pinned at its floor $1/2^{6}=0.0156$, so it reports one bit of sign agreement, not a magnitude; the paired
median difference carries no sealed bootstrap interval and we do not report it as an effect. The separation
therefore stands on the two per-edge intervals, not on the paired test. We state it as an
\emph{association}, not a cause. The natural day-to-day $4$-$5$ fidelity varies too little to serve as a
regressor (five daily values spanning $0.9444$--$0.9489$), so a slope fitted to those points would have
negligible leverage; and a good-versus-bad edge contrast confounds coupler fidelity with everything else
that differs between two physical couplers.

The causal upgrade is therefore not an observed correlate but a manipulation: we take the coupler fidelity
as a knob and set it ourselves. That experiment has now been run, and it is reported next
(Section~\ref{sec:rcausal}). It supersedes the earlier CZ-amplitude proof-of-control sweeps and delivers
the measured slope that this section deliberately does not claim from the observational contrast.

\subsection{A measured causal law: the coupler drive controls the certified entanglement}\label{sec:rcausal}

\begin{figure}[t]
\centering
\includegraphics[width=0.92\linewidth]{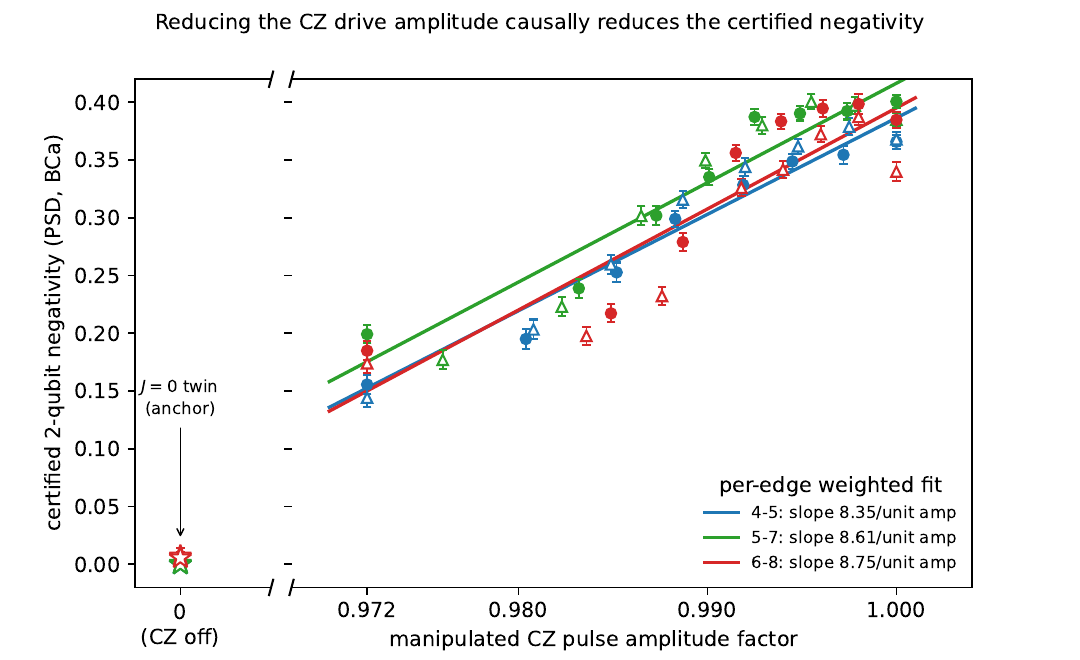}
\caption{\textbf{Reducing the manipulated CZ drive amplitude reduces the certified negativity along a
positive, well-fit slope over the probed ladder.} Certified two-qubit negativity (PSD, BCa) against the
\emph{manipulated} CZ pulse amplitude factor (the single knob we set), for three couplers of component $B$
($4$-$5$, $5$-$7$, $6$-$8$), two calibration sessions each (filled circles, session~1; open triangles,
session~2). Lines are the per-edge weighted fits over the rung window; the legend gives the per-edge slope
in certified negativity per unit fractional amplitude. The fits are \emph{local} linearisations: the
drive-to-phase map overshoots $\pi$ at the nominal amplitude and the certified negativity peaks at
$\lvert\zeta\rvert=\pi$, so the response is non-monotone across the nominal rung (in four of six runs the
negativity is higher one step below nominal), and the slope must not be extrapolated to zero amplitude. Open
stars are the CZ-off $J=0$ twins, a separate off-state control (zero amplitude, separable) and not points on
the fitted lines; the $x$-axis is broken between the twins and the rung window. The mean fringe visibility
(decoherence diagnostic) is held constant across the sweep, so the response is coherent. The independently
measured average gate fidelity $F_\mathrm{avg}$ is non-monotone in this drive (phase wrapping) and is
reported as a diagnostic in Appendix~\ref{app:B}, not used as the axis.}
\label{fig:causal}
\end{figure}

We now report the manipulation experiment, the paper's central device-physics result. On three physical
couplers of component $B$ ($4$-$5$, $5$-$7$, $6$-$8$), on two independent calibration days each, we
deliberately varied the CZ pulse drive amplitude of a fixed Bell preparation --- not of the QAOA cost layer
itself --- and certified the two-qubit negativity at each rung. The pre-registered decision rule is a
positive fitted slope whose interval excludes zero. All six sealed records carry a \texttt{GO} verdict, but
that verdict was evaluated on the sealed $F_\mathrm{avg}$ regression with a shot-level bootstrap that we
repudiate (Section~\ref{sec:deviations}, item~(e)); we therefore re-evaluate the gate, independently and
post-hoc, on the correct causal axis --- the manipulated drive (Fig.~\ref{fig:causal}) --- where it is again
met, and we do not present the sealed $F_\mathrm{avg}$-axis \texttt{GO} as corroborating the corrected-axis
result.

The right regressor is the drive we set, not the gate fidelity we measure. As the amplitude is swept the
coherent conditional phase passes through $\pi$ and wraps, so the average gate fidelity $F_\mathrm{avg}$ is
\emph{non-monotone} in the drive (it correlates with the certified negativity only weakly and
non-monotonically, Pearson $r=0.12$--$0.36$ across the six runs, and its coupling-off twin sits mid-axis at
$F_\mathrm{avg}\approx0.37$--$0.40$, not at zero); a straight-line slope on $F_\mathrm{avg}$ is therefore
uninterpretable and its shot-level interval understates a large rung-to-rung lack of fit (reduced
$\chi^2\approx100$--$137$). We accordingly regress the certified negativity on the manipulated CZ amplitude
factor, where the response is a positive, well-fit slope over the probed ladder. That slope is a
\emph{local} linearisation, not a global monotone law, and we say so explicitly. The manipulated amplitude
maps monotonically onto the conditional phase $\lvert\zeta\rvert$ only up to $\pi$; at the nominal amplitude
the phase already overshoots $\pi$ ($\lvert\zeta\rvert_\mathrm{nom}=3.3$--$3.6$~rad on all six runs), and
the certified negativity peaks at $\lvert\zeta\rvert=\pi$. The amplitude-to-negativity response is therefore
\emph{non-monotone across the nominal rung}: in four of the six runs the measured negativity is
\emph{higher} one amplitude step below nominal than at nominal (by $1.5$--$1.7$ per-rung bootstrap standard
deviations, and by $6.4$ on edge $6$-$8$ session~2, where it rises from $0.340$ to $0.387$). The fit is
valid only over the probed rung window near nominal; extrapolating its slope to zero amplitude is unphysical
(the amplitude-axis intercepts are large and negative), and the CZ-off twin (below) is a separate off-state
control, not a point on the line. The effect is causal, because the drive was set and
not observed. Per run the dose-response is a positive, well-fit slope ($R^2=0.83$--$0.96$), and with
rung-level lack-of-fit-honest standard errors every one of the six $95\%$ intervals excludes zero: the
per-run slopes are $8.1\,[6.6,9.7]$ and $8.7\,[6.6,10.8]$ on edge $4$-$5$, $8.1\,[5.7,10.6]$ and
$9.5\,[6.2,12.8]$ on edge $5$-$7$, and $9.0\,[5.2,12.7]$ and $8.5\,[4.7,12.4]$ on edge $6$-$8$, in units of
certified negativity per unit fractional CZ amplitude (equivalently $\approx0.085$ per one percent of
amplitude). Each one-sided slope test has $p<10^{-3}$, and the per-coupler tests survive both Holm and
Benjamini--Hochberg correction across the three couplers. Three couplers, six calibration runs, six
concordant positive slopes.

Two controls make the reading causal rather than incidental. The CZ-off rung, with the pulse amplitude set
to zero, is a pre-committed off-state control (not one of the fitted rungs): it returns a certified
negativity indistinguishable from zero on all six runs (PSD point estimates $0.000$--$0.007$, every BCa interval including zero), so with
the drive switched off there is no entanglement to reduce. And the effect is \emph{coherent}, not a loss of
coherence: while the drive was swept, the mean fringe visibility --- our separate decoherence diagnostic
--- stayed in a narrow band on each coupler ($\approx0.82$--$0.87$ on $4$-$5$, $\approx0.83$--$0.89$ on
$5$-$7$, $\approx0.76$--$0.81$ on $6$-$8$), and the CZ-off twins sat higher still ($\approx0.89$--$0.95$).
The negativity tracks the drive, not the visibility.

On the manipulated axis the pooled magnitude is also well behaved. Inverse-variance pooling of the two
rung-honest session slopes gives per-coupler slopes of $\mathbf{8.35}\,[7.33,9.36]$ ($4$-$5$),
$\mathbf{8.61}\,[7.04,10.17]$ ($5$-$7$) and $\mathbf{8.75}\,[6.60,10.91]$ ($6$-$8$), each excluding zero,
in the same units. Unlike the discarded $F_\mathrm{avg}$ regression --- on which the degraded coupler
spuriously appeared four times weaker --- the three couplers now agree: the between-coupler heterogeneity is
negligible ($Q=0.15$ on two degrees of freedom, $I^2=0\%$, $\tau^2=0$; with only $k=3$ couplers these
heterogeneity estimates are near-uninformative). The random-effects (DerSimonian--Laird) pool is
$\mathbf{8.47}\,[7.68,9.26]$; the conservative small-sample Knapp--Hartung--Sidik--Jonkman interval also
excludes zero once its variance-inflation factor is floored to one ($[6.73,10.21]$). We floor it because the
raw factor $q=0.076<1$ reflects underdispersion at $k=3$, so an unfloored KHSJ interval (the
anti-conservative $[7.99,8.95]$) would understate rather than widen the small-sample uncertainty. The causal
claim rests on the per-coupler evidence --- three of three couplers, six of six runs, every rung-honest
interval excluding zero, Holm- and BH-robust --- and the pooled magnitude ($\approx0.085$ of certified
negativity per one percent of CZ amplitude) is consistent with it. The full per-rung tables with $R^2$ and
lack-of-fit, the $F_\mathrm{avg}$ diagnostic, the equivalent $\lvert\zeta\rvert$-axis fit, the per-session
provenance and the external timestamps are in Appendix~\ref{app:B}.

\subsection{The certificate collapses with depth on the primary instance}\label{sec:r4}

The entanglement present at depth one is gone by depth three on the same instance. On the good edge of
\texttt{b08} the raw negativity is $\mathbf{0.000}$ at $p=3$ on all six days (grand mean $0.000$,
between-day SD $0.000$, mitigated $0.000$), against $0.077$ at $p=1$ (Fig.~\ref{fig:ceiling}, left); the
$p=3$ arm is TOST-equivalent to zero at margin $0.02$ (upper $95\%$ bound $0.000$, all six days exactly
zero), like the other controls. This is a within-instance depth collapse ($N=1$): the same edge, the same
instance, a single depth step. The loss is not merely a hardware artefact of stacking gates. Even the
noiseless target degrades: the ideal two-qubit negativity on this cut halves from $0.3196$ at $p=1$ to
$0.1728$ at $p=3$, while the measured hardware value falls all the way to zero, faster than the ideal
declines. We claim no more than this within-instance collapse. In particular, the depth-three ensemble
null of Section~\ref{sec:r3} was measured at one depth only; it speaks to the absence of an optimisation
signal \emph{at} $p=3$, is not evidence of a decline across depth, and we do not use it to corroborate the
collapse. We do not claim a mediation-tested shared cause. The pre-registered out-of-sample noise-model
test --- fit at $p=1$, predict $p=2,3$ --- requires the depth-one-and-two ensemble and is deferred.

\subsection{No optimisation signal at depth three: a pre-registered honest negative}\label{sec:r3}

\begin{figure}[t]
\centering
\includegraphics[width=\linewidth]{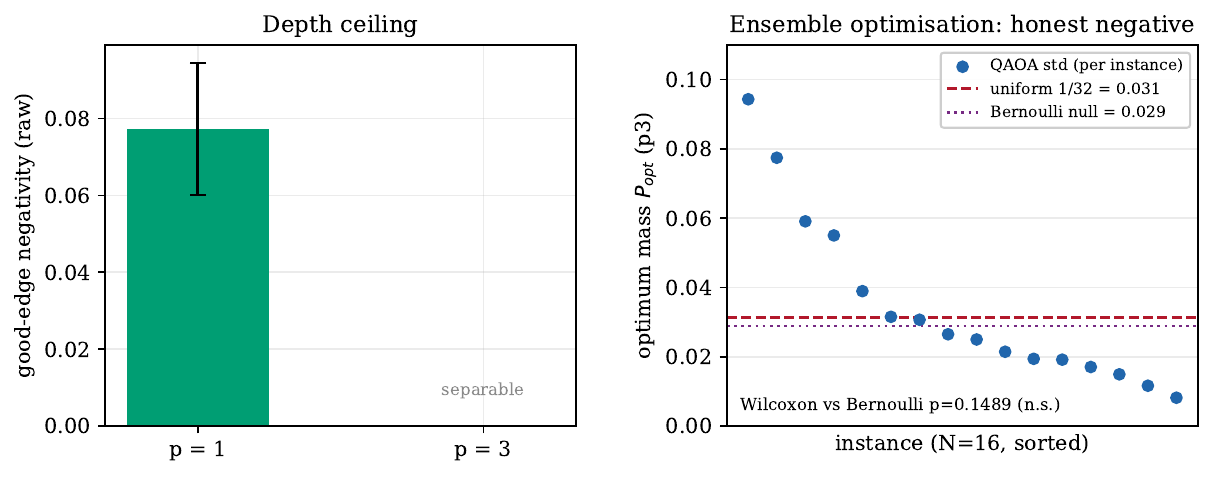}
\caption{\textbf{The certificate collapses by depth three on \texttt{b08}; separately, the depth-three
ensemble shows no optimisation signal.} Left: the good-edge negativity present at $p=1$ is separable by
$p=3$ (instance \texttt{b08}, day-level over six calibrations) --- a within-instance depth collapse. Right:
across the $N=16$ ensemble measured at $p=3$ only, per-instance optimum mass $P_\mathrm{opt}$ sits at the
uniform ($1/32$) and device-marginal Bernoulli floors (one-sided Wilcoxon versus Bernoulli $p=0.15$, not
significant); this speaks to $p=3$ alone and is not evidence of a decline with depth.}
\label{fig:ceiling}
\end{figure}

We report the primary optimum-mass result up front, because it is evidence that bears against a naive
optimisation reading and we do not bury it. Across the pre-registered $N=16$ instance ensemble at $p=3$,
the probability mass on the unique optimum sits at the device floor: the standard-arm grand-mean
$P_\mathrm{opt}$ is $\mathbf{0.034}$ against a device-marginal Bernoulli null of $0.029$ and a uniform
floor of $0.031$, with $10/16$ instances above the null and a one-sided Wilcoxon $\mathbf{p=0.15}$; the
warm-start arm gives $0.033$ against $0.032$, $p=0.28$. The mean approximation ratio is consistent with
zero ($r\approx0.01$--$0.02$). No optimisation signal is detected at $p=3$ (Fig.~\ref{fig:ceiling},
right). This is a pre-registered honest negative, and it is under-powered by design: $p=3$ is exactly the
depth at which the certificate has already collapsed. It must not be read as depth-general. On the single
primary instance \texttt{b08} at $p=1$, the coupled arm places $0.057$ of its mass on the optimum against
$0.004$ for the $J=0$ twin, a coupled-versus-off difference of $0.054$ driven by the coupling rather than
by single-qubit bias. This coupled advantage is present at $p=1$ but does
\emph{not} decline monotonically with depth on \texttt{b08} ($0.054$ at $p=1$, $0.013$ at $p=2$, $0.026$
at $p=3$; non-monotonic), so we make no ``shrinks with depth'' claim from it. It is single-instance and
does not rescue the ensemble-level null: whatever coupled optimisation signal exists at low depth on this
one instance does not survive to an ensemble-level effect at $p=3$. The depth-one-and-two ensemble and the
scrambled-encoding null are staged for a future allocation.

\subsection{An independent witness, consistent on Red}\label{sec:r1b}

The randomized-measurement $p_3$-PPT witness gives an independent view of the same claim, reported
honestly for what it is. On Red, on the good edge at $p=1$, the witness is
$\mathbf{W=p_2^2-p_3=0.0105}$ over $N_U=800$ settings (measured moments $p_2=0.407$, $p_3=0.155$), with a
BCa $95\%$ CI $\mathbf{[0.0060,0.0161]}$ excluding zero and lying $7.2$ null standard deviations above the
Werner PPT-boundary null (threshold $t_{95}=0.0027$; empirical false-positive rate of the CI-excludes-zero
rule $1.5\%$). Against the Werner boundary the signal is unambiguous. But at this small magnitude the
honest near-pure-separable null is not controllable at a feasible number of settings, as the power study
shows (it would need $N_U\approx12{,}672$ against the $800$ run), so on Red we report the witness as
\emph{consistent with entanglement}, not as a certification. Measured on a later calibration than the
six-day tomography, it is in any case a cross-method reproduction of the entanglement claim on a nominally
identical preparation, not a second reading of the $0.077$ negativity; mapping $W$ to a negativity is
model-dependent and we do not perform it. The same estimator is fully powered where the realised
entanglement is large, which we treat as a scope note rather than a result (Section~\ref{sec:limits}).

\subsection{Six-day stability, warm-start and readout mitigation}\label{sec:r7}

The certified negativity is reproducible across the recalibration cycle. Between-day standard deviation is
$0.016$ over the six calibrations, with a day-level intraclass correlation of $0.85$; we report this as a
preliminary stability characterisation at the recalibration cadence, not as a drift bound, which would
need daily-or-finer sampling over weeks to months. Warm-start also certifies entanglement at $p=1$:
day-level negativity $0.057$ (hierarchical CI $[0.048,0.066]$, excluding zero on $6/6$ days), below the
cold-start value, consistent with the warm mixer's reduced coherent evolution; at $p=3$ warm and cold are
indistinguishable, both at the floor. Readout mitigation (IBU) raises the good-edge negativity from
$0.077$ to $0.152$ and is reported alongside the raw value throughout, with the raw value as the
certificate; the mitigated twin residual stays an order of magnitude below the certificate
($0.015$ versus $0.152$). The ZNE-folding rung was not run in this reduced pilot and is deferred. Every
pre-registered outcome is reported, including the honest negatives, with no metric-switching and no
cherry-picking.

\begin{figure}[t]
\centering
\includegraphics[width=0.86\linewidth]{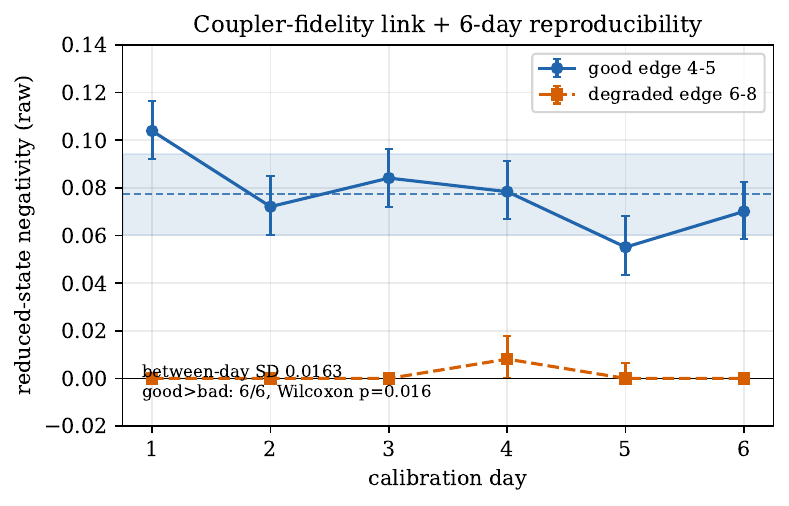}
\caption{\textbf{The good coupler certifies on every calibration; the degraded coupler never does.} Raw
two-qubit negativity of the good edge $4$-$5$ (upper) and the degraded edge $6$-$8$ (lower) across the six
daily calibrations, with the good-edge grand-mean band. Good above bad on $6/6$ days; the separation is
reported through the two per-edge day-level intervals (good BCa CI $[0.065,0.091]$, degraded
TOST-equivalent to zero), because the pre-registered paired test is a secondary hypothesis that does not
survive multiplicity correction (see text). Between-day standard deviation $0.016$.}
\label{fig:coupler}
\end{figure}

\section{Discussion}\label{sec:discussion}

The result of record is narrow and firm: a corrected QAOA cost layer, run on the good coupler of a
nine-qubit superconducting processor at depth one, prepares a two-qubit state whose bipartite
entanglement is certified on hardware, reproducibly across six calibrations, against a separable on-device
control. The negativity is $0.077$ with a BCa interval $[0.065,0.091]$ that excludes zero on every day.
That number is small --- roughly a quarter of the ideal $0.3196$ --- and it should be: the good CZ gate
runs at $94.7\%$, and the depth-one circuit sits well below its noiseless target. What matters is that the
certificate is a measured property of the device, protected on one side by a compilation guardrail that
proves the compiled circuit encodes the intended coupling rather than the diagonal fault, and on the other
by a coupling-off twin that certifies zero on the same hardware --- the twin, not the guardrail, is what
shows the coupling is realised on the device. This is the answer to the standard circularity objection
against hardware entanglement claims: the entanglement is neither written into the generator nor
manufactured by mitigation.

The device-physics reading has two parts, and we keep them logically distinct. The first is
observational: on the same chip, in the same instance, the $94.7\%$ coupler certifies and the $74.7\%$
coupler does not, with the good edge ahead on every day. We report this as an association, not a cause,
because a good-versus-bad edge contrast covaries coupler fidelity with everything else that differs
between two physical couplers, and the natural day-to-day fidelity range is too narrow to fit a response;
the pre-registered paired test of the ordering is a secondary hypothesis that does not survive
multiplicity correction, so the separation itself rests on the two per-edge day-level intervals. The
second part is the causal statement, and here we do have it, because we manipulated the coupling rather
than observing it. The pre-registered experiment shows directly that a coupler's two-qubit drive
\emph{controls} the entanglement it can certify: across three couplers and two calibration days, reducing
the CZ pulse amplitude lowers the certified negativity along a positive, well-fit slope over the probed
ladder ($R^2=0.83$--$0.96$, every per-run rung-honest interval excluding zero, Holm- and BH-robust), with
the CZ-off twin a separate off-state control (not a fitted point) and the fringe visibility held constant so
the mechanism is coherent. The fitted slope is a local linearisation of a peaked response --- the certified
negativity peaks at conditional phase $\lvert\zeta\rvert=\pi$, which the nominal amplitude slightly
overshoots, so the response is non-monotone across the nominal rung --- and we read it as a positive
dose-response over the probed window, not as a claim of global monotonicity or of behaviour at zero drive.
We are deliberately careful about what is and is not shown. The clean causal variable is the
manipulated drive itself; the average gate fidelity $F_\mathrm{avg}$ is non-monotone in that drive because
its conditional phase wraps, so we do \emph{not} claim that coherent gate fidelity governs entanglement,
only that the two-qubit drive does. This is the sense in which the coupling, not qubit count, sets the
certified entanglement on this device: a measured dose-response, not an inference from the good-versus-bad
contrast. What our data do not yet support is the transfer of this law off a single device or its slope on
the QAOA cost layer itself rather than a Bell preparation; we mark that line clearly
(Section~\ref{sec:limits}).

The depth collapse positions the work against the QAOA-hardware literature. Deeper circuits should, and do,
lose the signal as they add entangling gates the device cannot sustain~\cite{stilckfranca2021limitations,wang2021noiseinduced}.
On this instance even the noiseless target does not protect the certificate: the ideal two-qubit negativity
on the good cut itself halves from $p=1$ to $p=3$ (from $0.3196$ to $0.1728$) while the hardware value
collapses to zero, faster than the ideal declines~\cite{harrigan2021qaoa,weidenfeller2022scaling}. We
report the collapse as a within-instance observation on \texttt{b08} ($N=1$), and we do not assert a
mediation-tested cause; the out-of-sample noise model that would test the mechanism directly is deferred.
Kept deliberately separate is the depth-three ensemble null, a pre-registered honest negative reported up
front, not a buried inconvenience: across the ensemble at the one depth we measured there is no
optimisation signal. That null was taken at a single depth; we do not read it as corroborating the
collapse, and it establishes only that no optimisation signal is present at $p=3$. On error mitigation our stance is
conservative by design. Readout unfolding raises the negativity but we never let it carry the claim, and
we did not run zero-noise extrapolation, whose artefactual gains on hardware are now
documented~\cite{koster2026zne}. The economical reading of the whole picture is that value on this device
comes from realising the coupling correctly on a healthy edge at low depth, not from deeper circuits or
post-hoc extrapolation. The independent randomized-measurement route reinforces the certificate on Red,
where it is consistent with entanglement, and its power behaviour is the subject of a general lesson below.

The methodological posture is where this study departs from much of the hardware QAOA
literature~\cite{harrigan2021qaoa,weidenfeller2022scaling,shaydulin2024scaling}. That literature reports
success probabilities and approximation ratios; the field routinely quotes standard deviations, box plots
or shot-noise bars, but rarely formal confidence intervals, and it rarely proves that the two-qubit
coupling is realised on the device at all. We invert those priorities. Every headline carries a
bias-corrected bootstrap interval at a named unit; the compilation is proven correct by a per-circuit
statevector guardrail before a single shot is spent; the entanglement is read from on-device tomography
against a separable $J=0$ twin; and the honest negatives are stated in the abstract, not the appendix. This is deliberately the
answer to the standard structural objections a sceptical referee raises against near-term hardware
claims --- circularity, missing baselines, weak statistics, and conclusions asserted beyond the evidence.
The cost is scope: a single device, a single certified instance, and an expensive measurement budget. We
regard that trade as the correct one for a certification paper, whose value is in what it proves rather
than in how far it reaches.

The randomized-measurement case also carries a general lesson for entanglement certification on
near-term hardware~\cite{elben2020ppt,neven2021ptmoments,elben2022toolbox}. Moment witnesses are
attractive because they avoid full tomography, but their variance grows sharply for high-purity separable
states, so at small realised entanglement the honest separable null cannot be controlled at feasible
sampling. A witness that is many standard deviations clear of a Werner boundary null can still be
underpowered against the null that actually threatens the claim. We think this distinction --- between the
convenient null and the threatening null --- deserves to be standard practice, and we make it load-bearing
here by downgrading the on-Red witness accordingly while retaining it where its magnitude makes it sound.

\section{Limitations}\label{sec:limits}

This is a single-device study, and every conclusion is bounded accordingly. The certificate is
single-instance (\texttt{b08}, chosen by a pre-registered entropy criterion) on one edge of one processor;
whether it transfers to other instances, edges, devices or fidelity regimes is untested here. The causal
dose-response is genuine but its scope is specific: it is measured on a Bell preparation, not on the QAOA
cost layer, on three couplers of one processor across two calibration days, so it establishes that the
manipulated CZ drive controls certified negativity \emph{on this device} and not that the same slope holds
on the QAOA layer or on other hardware. The reported slope is moreover a local linearisation over the probed
amplitude window near nominal, where the drive-to-phase map is monotone up to $\pi$; because the certified
negativity peaks at $\lvert\zeta\rvert=\pi$ (which the nominal amplitude overshoots), it is a positive
dose-response over that window, not a global monotone law, and must not be extrapolated to zero drive (the
$J=0$ twin is a separate off-state control, not a fitted point). We regress on the manipulated drive rather than on the measured
average gate fidelity $F_\mathrm{avg}$, because $F_\mathrm{avg}$ is non-monotone in the drive (its Ramsey
conditional phase wraps); we therefore do not claim that coherent gate fidelity governs entanglement, only
that the two-qubit drive does. On the manipulated axis the three couplers are statistically homogeneous
($I^2=0\%$), so the pooled magnitude is well defined; we nonetheless carry the causal claim primarily on the
per-coupler intervals, and with only $k=3$ couplers the pool remains a summary rather than an independent
test.
The depth collapse is a within-instance observation on \texttt{b08} ($N=1$), not a mediation test; the
out-of-sample noise model (fit at $p=1$, predict $p=2,3$) needs the depth-one-and-two ensemble and is
deferred, as is the scrambled-encoding hardware null for the optimum-mass test, which is therefore only
partially tested. The depth-three ensemble null is measured at one depth only and bears on $p=3$ alone, not
on any decline with depth. The randomized-measurement witness is underpowered on Red at the measured
magnitude and is reported there only as consistent with entanglement; for completeness we note the same
witness clears its threshold on the $53$-qubit superconducting Euro-Q-Exa device (EuroHPC award
EHPC-QCP-2026Q01-122) in a single session where the realised magnitude is larger ($W=0.1067$, BCa CI
$[0.0865,0.1279]$), which we report as a non-load-bearing cross-architecture corroboration, not a
day-replicated certification. The claim is bipartite across the named good-edge cut only; we make no
genuine-multipartite claim without a violated multipartite witness. The measurement cost is high and
worth stating plainly: $36{,}864$ tomography shots per arm per day over six days, plus a $320{,}000$-shot
randomized-measurement budget and the eight-rung causal ladders, for a single instance and three couplers.
Finally, the instances are classically trivial; the value is methodological, and no advantage, speedup,
scaling or solver claim is made.

\section{Conclusions}\label{sec:conclusion}

On this device, the corrected QAOA cost layer prepares at depth one, on the good coupler, a two-qubit state
whose bipartite entanglement is certified by a tomographic negativity of $0.077$ (BCa $95\%$ CI
$[0.065,0.091]$, excluding zero on all six calibrations); a per-circuit statevector guardrail proves that
the compiled layer encodes the intended coupling rather than the diagonal fault, the coupling-off $J=0$
twin returns separable ($0.00$, TOST-equivalent to zero) on the same hardware, and the degraded coupler
does not certify, so the certified entanglement is neither written into the generator nor manufactured by
mitigation. Beyond certifying that entanglement, we measure what controls it: in a pre-registered
manipulation experiment on three couplers over two calibration days, on a fixed Bell preparation rather
than the QAOA cost layer, deliberately reducing the CZ pulse drive amplitude causally reduces the certified
negativity along a positive, well-fit slope over the probed amplitude ladder ($R^2=0.83$--$0.96$), every
one of the six rung-level (lack-of-fit-honest) $95\%$ intervals excluding zero (one-sided $p<10^{-3}$,
Holm- and BH-robust across the three couplers, which agree to within their intervals). This slope is a
local linearisation of a peaked response --- the certified negativity peaks at conditional phase
$\lvert\zeta\rvert=\pi$, which the nominal amplitude overshoots, so the response is non-monotone across the
nominal rung --- and the CZ-off $J=0$ twin is a separate off-state control, not a point on the fitted line.
The two-qubit drive, not qubit count, sets the certifiable entanglement on this device, measured directly
rather than inferred from a within-chip contrast; the average gate fidelity is non-monotone in the drive
and is reported only as a diagnostic. On the primary instance the certificate collapses by depth three;
separately, the pre-registered $N=16$ ensemble measured at depth three shows no optimisation signal
(one-sided Wilcoxon $p=0.15$): a pre-registered honest negative, reported up front, that speaks to depth
three alone, and the independent randomized-measurement witness is underpowered on this device at the
realised magnitude and is reported there only as consistent with entanglement.

The contributions of this study are methodological. We certify bipartite entanglement in a corrected QAOA
cost layer on real hardware and, through the guardrail-and-twin construction, make that certificate proof
against the standard circularity objection; we turn the account from observational to causal by
pre-registered manipulation of the two-qubit drive; we hold every headline to a bias-corrected confidence
interval at a named inferential unit, where the hardware-QAOA literature reports success probabilities and
approximation ratios but rarely formal intervals, and we bind every reported number to its raw counts, seed
and in-job guardrail pass by a SHA-256 deposit ($121/121$ checksums verified) that can be re-checked rather
than trusted; and we report the honest negatives at the front, drawing from the underpowered witness a
transferable lesson for near-term certification --- a moment witness many standard deviations clear of a
convenient Werner-boundary null can still be underpowered against the separable null that actually
threatens the claim, so the threatening null, not the convenient one, must set the sampling budget.

The concrete next steps are three: carry the causal dose-response onto the QAOA cost layer itself and onto
further devices to test its transfer; measure the depth-one-and-two ensemble to test the depth mechanism
out-of-sample; and repeat the certificate on further instances and edges. We claim no advantage: the
five-variable tree Ising instances are classically trivial, and the contribution is a rigorous, reproducible
and causally grounded account of what one real device does to a corrected QAOA cost layer and of how its
coupler's two-qubit drive controls the entanglement it certifies.

\section*{Data and code availability}\label{sec:data}
The reproducibility deposit covers the primary-device (Red) data: the sealed pre-registration, all raw
per-shot outputs and per-job provenance for the six-day tomography (2026-08-06..11, jobs 11479--11484) and
the Red randomized-measurement certificate (2026-08-13, jobs 11608--11611), the exact instances and frozen
angles, the device runcard snapshots, and the full analysis and figure pipeline that regenerates every
reported Red number and figure; no new hardware run is needed to reproduce any reported Red quantity. The
cross-architecture Euro-Q-Exa randomized-measurement run (captured 2026-08-20 under EuroHPC award
EHPC-QCP-2026Q01-122) is held under the hosting facility's data policy and is \emph{outside} this deposit;
the single certified quantity there ($W=0.1067$, BCa CI $[0.0865,0.1279]$) is a corroboration, not part of
the primary certificate. A single SHA-256 manifest binds every Red figure and number to its raw counts,
analysis script, instance seed, runcard-snapshot hash and in-job guardrail pass (deposit integrity:
121/121 checksums verified). The master seed is \texttt{20260804}; the executed component-B instance set
\texttt{instances\_liveB.json} has SHA-256
\texttt{20bba8250662f3976996aee00a13a8d183e8c1074a77f35390dc0e371a27e309} (this is the hash of the
deposited, executed file; the earlier live-B re-seal recorded \texttt{e78a9b10\ldots} for the same
instance set, an amendment chain documented in the deposit's reconciliation note), the primary instance
record \texttt{b08} has SHA-256 \texttt{ba4aea8f\ldots}, and the sealed pre-registration
\texttt{REGISTRATION.md} has SHA-256
\texttt{84f2b23699e37ac84e88fdf8b7438aa2fdff2bd4beb63bef8d3d62af54ef575a}.

The causal manipulation experiment (Sections~\ref{sec:causalmethods}, \ref{sec:rcausal}) comprises six
sealed per-session pre-registrations, six raw result records with per-rung tomography and Ramsey data, the
per-job and master runcard snapshots with their SHA-256, and the re-analysis script that refits the
dose-response on the manipulated CZ amplitude (equivalently on the physically unwrapped $\lvert\zeta\rvert$)
with rung-level lack-of-fit-honest errors from the per-rung arrays in each record. These six
\texttt{results.json}, their RFC-3161 timestamp tokens (\texttt{.tsr}) and the re-analysis script
\texttt{causal\_reanalysis.py} are present in the project working tree and are provided to reviewers as a supplementary archive under
\texttt{Tests/piloto\_5q\_20260726/eurohpc/}; they will be folded into the Zenodo bundle, with the manifest
checksums regenerated to include them, at the coordinated re-publish. The complete deposit (Red component-B plus
the causal campaign) is archived on Zenodo under the reserved DOI \texttt{10.5281/zenodo.22262058}; the record
is a private draft during review and its public DOI is minted at acceptance. The script reproduces every per-run amplitude-axis slope, $R^2$ and reduced $\chi^2$
of Table~\ref{tab:causal}, the $F_\mathrm{avg}$ diagnostic and unwrapped $\lvert\zeta\rvert$ columns of
Table~\ref{tab:causaldiag}, and the pooled slopes of Table~\ref{tab:causalpool} from the sealed records
(self-test), and supersedes the earlier $F_\mathrm{avg}$-axis \texttt{pool\_slopes.py}. Each per-session
pre-registration was
sealed and independently timestamped by an RFC-3161
Time-Stamp Authority (freetsa.org) \emph{before} that day's data existed: the three session-2
pre-registrations were granted at 2026-09-02 01:54\,Z and the three session-1 pre-registrations at
2026-09-01 08:23\,Z, and each token's message imprint equals the SHA-256 of its pre-registration file
(for example, the edge-$4$-$5$ session-2 token imprint
\texttt{e778820ffa015e721c9e4a658b773bd1bea4790be66d57c2377fb48a2d950f2e} matches that file). The six
pre-registration SHA-256 values are \texttt{a11a260e\ldots} (4-5 s1), \texttt{e778820f\ldots} (4-5 s2),
\texttt{59cc9177\ldots} (5-7 s1), \texttt{1279eb06\ldots} (5-7 s2), \texttt{96554e7d\ldots} (6-8 s1) and
\texttt{85155616\ldots} (6-8 s2); the session master runcards hash \texttt{e8db3ca5\ldots} (s1) and
\texttt{d597a096\ldots} (s2). Data and figures are released under CC-BY-4.0 and code under the MIT licence.

\section*{Author contributions}
C.-M.L.B.\ conceived and directed the study, designed the pre-registered protocol, executed the hardware
campaign, performed the analysis, and wrote the manuscript.

\section*{Use of AI tools}
Large-language-model-based tools assisted with data-analysis and figure-generation code, with
bibliographic search and formatting, and with drafting and language editing of the manuscript. All
scientific claims, all data and their interpretation were verified by the author, who takes full
responsibility for the content.

\section*{Competing interests}
The author declares no competing interests.

\section*{Acknowledgements}
This work was supported by grant INNO-2026-1-0007. We thank the Barcelona Supercomputing Center for access
to the Red superconducting processor on MareNostrum Ona; the cross-architecture randomized-measurement run
used the Euro-Q-Exa processor under EuroHPC award EHPC-QCP-2026Q01-122. Hardware-access acknowledgements
follow each facility's policy.

\appendix
\section{Pre-registration}\label{app:prereg}
The analysis plan, instance set, frozen angles, hypotheses, nulls and forbidden-claims list were sealed
(SHA-256 fixed) before any hardware execution; the sealed \texttt{REGISTRATION.md} has SHA-256
\texttt{84f2b236\ldots} (above). We quote the sealed primary hypothesis as written and then disclose our
departure. \emph{Primary hypotheses, as sealed.} PP1 (entanglement) names the \emph{$p_3$-PPT
randomized-measurement (RM) moment witness} as the primary certificate --- ``the $p_3$-PPT mixed-state
moment witness certifies bipartite entanglement across the certified cut $\ldots$ cross-checked by
two-qubit reduced-state negativity'' --- so the sealed plan makes the RM witness primary and the tomographic
negativity its cross-check, while the coupling-off $J=0$ twin and the $\gamma=0$ controls must not certify
(TOST equivalence). PP2 (optimum-mass): across $N=16$ instances, QAOA places more probability on the unique
optimum than the spectrum-preserving scrambled-encoding null and the device-marginal Bernoulli null.
\emph{Endpoint elevation (disclosed departure).} We invert PP1's roles and certify on the negativity, with
the RM witness reported as consistent-with-entanglement, because the RM witness is underpowered at the
magnitude realised on Red: pushing the near-pure-separable null threshold below the measured signal needs
$N_U\approx12{,}672$ settings against the $800$ run. This is a \emph{principled post-hoc} elevation on a
data-independent power ground (the power calculation depends only on the null and the target magnitude, not
on the RM outcome), not a data-dependent (HARKing) switch; the negativity certificate carries its own
sealed BCa interval whose validity does not depend on the RM result. We make no claim that the power study
was sealed before unblinding, and the sealed RM-primary plan is reported faithfully here; the full account
is in the deposit's amendment note (\texttt{AMENDMENTS.md}, section~3). \emph{Secondaries
(Holm/BH-controlled):} H3 entanglement grows $p{=}1\to p{=}2$; H4 good edge $>$ degraded edge; H5 warm $>$
cold; H6 readout mitigation helps and ZNE-folding does not; H7 depth ceiling; H8 day-level drift variance
over $\geq6$ days; H9 a CZ-anchored noise model predicts $p=2,3$ out-of-sample from a $p=1$ fit.
\emph{Nulls and statistics:} inferential-unit hierarchy shot $<$ run $<$ day $<$ instance
(randomized-measurement unit = setting); hierarchical BCa bootstrap for the certificate and its controls,
with a CI at a named unit and effect sizes throughout; scrambled-encoding and device-marginal Bernoulli
nulls, with uniform and random-angle floors; day-level stopping rule at CI half-width $\leq0.02$.
\emph{Forbidden claims:} no advantage, speedup, solver, supremacy, scaling or hardness; no withdrawn-era
number or old angles as ``optimal''; no entanglement from structure or simulator alone (the on-hardware
certificate with the $J=0$ control separable is required); no purity-as-entanglement; no
genuine-multipartite claim without a violated multipartite witness; the ideal $0.3196$ labelled ideal,
never hardware; association-not-causation for the \emph{observational} within-chip good-versus-degraded
edge contrast (the causal claim is licensed only by the separate, independently pre-registered CZ-amplitude
\emph{manipulation} experiment, Appendix~\ref{app:B}); the randomized-measurement witness never stated as a
certification at the measured on-Red magnitude; every hardware number tied to its runcard-snapshot hash,
in-job guardrail pass and daily calibration hash.

\section{Angles, instances and reconstruction}\label{app:angles}
The frozen depth-one angles for the primary instance \texttt{b08} are $\gamma=1.909173$,
$\beta=0.733634$, from exact $32$-state multi-start differential evolution with the correct
CNOT-conjugated ZZ (bounds $\gamma\in[0,4\pi)$, $\beta\in[0,\pi)$). The instance fields and couplings are
listed in Table~\ref{tab:b08}. The tomographic reconstruction uses nine Pauli-pair settings at $4096$
shots each; the negativity is evaluated on the nearest physical (PSD, unit-trace) reconstruction, with a
BCa bootstrap (per-day $B=8000$, hierarchical $B=20000$) whose acceleration comes from a
delete-one-shot-block jackknife. The randomized-measurement witness uses $N_U=800$ single-qubit-Clifford
settings at $N_M=400$ shots, with $p_2$, $p_2^2$ and $p_3$ estimated by distinct-index U-statistics and a
threshold calibrated at the $95$th percentile of $\hat W$ under a Werner PPT-boundary null.

\begin{table}[t]
\centering
\caption{Primary instance \texttt{b08} (component $B$, physical qubits $\{4,5,6,7,8\}$; local indices
$0$--$4$). Fields $h$ and native-edge couplings $J$; unique ground state $\texttt{10110}$, spectral gap
$0.68$; frozen depth-one angles $\gamma=1.909$, $\beta=0.734$. Instance record SHA-256
\texttt{ba4aea8f\ldots}.}
\label{tab:b08}
\begin{tabular}{lccccc}
\toprule
local index & $0$ ($q_4$) & $1$ ($q_5$) & $2$ ($q_6$) & $3$ ($q_7$) & $4$ ($q_8$) \\
\midrule
field $h$ & $0.984$ & $0.193$ & $0.344$ & $0.947$ & $-0.613$ \\
\midrule
edge & $4$-$5$ & $4$-$6$ & $5$-$7$ & $6$-$8$ & \\
coupling $J$ & $0.360$ & $-0.148$ & $0.171$ & $0.895$ & \\
\bottomrule
\end{tabular}
\end{table}

\begin{table}[t]
\centering
\caption{Ideal (noiseless statevector) versus measured (hardware) two-qubit negativity for the primary
instance \texttt{b08}, good edge $4$-$5$. The ideal target improves or persists with depth while the
hardware certificate collapses; the ideal $0.3196$ is labelled ideal and is never a hardware value. Also
shown are the partial-transpose moments $(p_2,p_3,W)$: ideal, measured on Red (underpowered at this
magnitude), and the single-session cross-architecture corroboration on the $53$-qubit Euro-Q-Exa device
(non-load-bearing).}
\label{tab:idealmeas}
\begin{tabular}{lccc}
\toprule
quantity & ideal & measured (Red) & measured (second arch.) \\
\midrule
negativity, edge $4$-$5$, $p{=}1$ & $0.3196$ & $0.077\ [0.065,0.091]$ & --- \\
negativity, edge $4$-$5$, $p{=}3$ & $0.1728$ & $0.000$ & --- \\
negativity, edge $6$-$8$, $p{=}1$ & $0.0922$ & $0.001\ [0.000,0.007]$ & --- \\
PT moment $p_2$ & $0.699$ & $0.407$ & --- \\
PT moment $p_3$ & $0.243$ & $0.155$ & --- \\
witness $W=p_2^2-p_3$ & $0.245$ & $0.0105\ [0.006,0.016]$ & $0.1067\ [0.087,0.128]$ \\
\bottomrule
\end{tabular}
\end{table}

\begin{table}[t]
\centering
\caption{Run-date device conditions of component $B$ from BSC daily autocalibration, six-day campaign
means ($T_2$ archived in ns, confirmed by BSC). These contextualise device health and are not used as
day-level regressors.}
\label{tab:device}
\begin{tabular}{lcc}
\toprule
quantity & good edge $4$-$5$ & degraded edge $6$-$8$ \\
\midrule
CZ-gate fidelity (campaign mean) & $94.7\%$ & $74.7\%$ \\
CZ-gate fidelity (RM day, 13 Aug) & $94.9\%$ & $74.0\%$ \\
single-shot readout & $\approx96\%$ & --- \\
$T_1$ ($q_4/q_5$) & \multicolumn{2}{c}{$\approx 22/32\,\mu$s} \\
$T_2$ ($q_4/q_5$) & \multicolumn{2}{c}{$\approx 20/7\,\mu$s} \\
\bottomrule
\end{tabular}
\end{table}

\begin{figure}[t]
\centering
\includegraphics[width=0.82\linewidth]{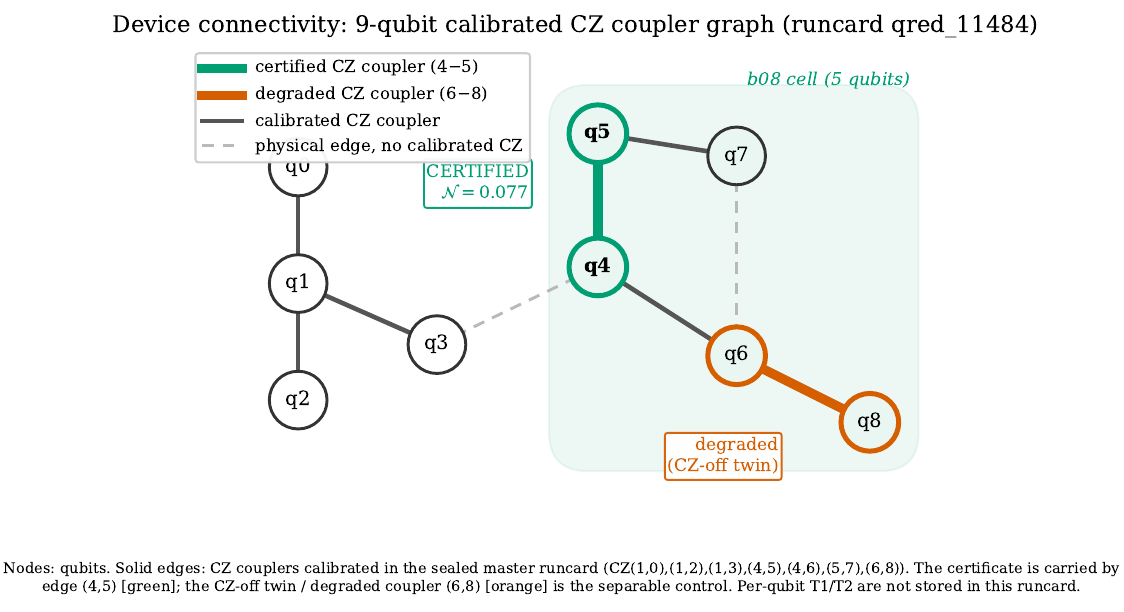}
\caption{\textbf{Component $B$ of Red and its couplers.} The calibrated connectivity read live from the
runcard on 4~August 2026: nine qubits, seven CZ couplers, and the absent $3$-$4$ bridge that splits the
chip into components $A=\{0,1,2,3\}$ and $B=\{4,5,6,7,8\}$. Component $B$ carries the primary instance
\texttt{b08}, the good edge $4$-$5$, the degraded edge $6$-$8$ and the $5$-$7$ coupler used in the causal
experiment.}
\label{fig:device}
\end{figure}

\begin{figure}[t]
\centering
\includegraphics[width=0.9\linewidth]{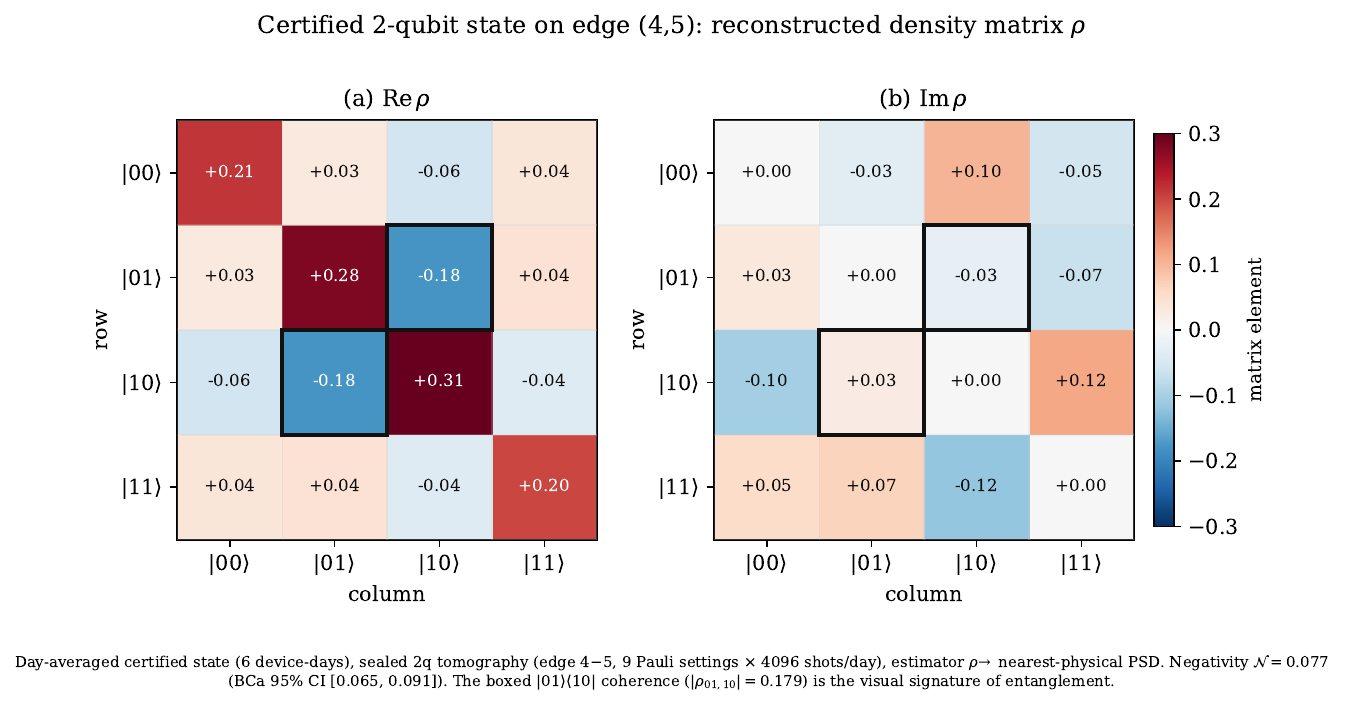}
\caption{\textbf{Reconstructed two-qubit density matrix on the good edge $4$-$5$ at depth one.} The nearest
physical (PSD, unit-trace) reconstruction from nine-Pauli-pair tomography on the primary instance
\texttt{b08}, from which the certified negativity of $0.077$ is read. The off-diagonal coherences carry the
entanglement; the coupling-off $J=0$ twin (not shown) reconstructs to a separable matrix on the same
device and readout.}
\label{fig:rho}
\end{figure}

A compiled-circuit schematic (the corrected CNOT-conjugated cost layer, the two-CZ budget, the mixer and
the statevector guardrail, set against the withdrawn diagonal fault) is described in
Section~\ref{sec:bug}; the drawn asset is not yet generated and is marked as a figure to add before
submission.

\section{Ablations, power, calibration and the causal experiment}\label{app:B}

This appendix collects the supporting evidence behind the main-text numbers: the arm-by-arm ablation with
its intervals, the randomized-measurement power and stopping-rule derivations, the per-day calibration and
readout characterisation, the record of withdrawn and deferred runs, and the full per-run tables and
external timestamps of the causal experiment. Every number is drawn from a sealed deposit file; where a
number was not located in a sealed file it is marked \texttt{TODO-VERIFY}.

\paragraph{Ablation across arms.} Table~\ref{tab:ablation} gives the day-level two-qubit negativity for
each pre-registered arm on the primary instance \texttt{b08}, with the certificate and its controls
carrying BCa intervals and the secondary tomographic monotones carrying day-level intervals. The pattern
is the one a certificate requires: the coupled good edge certifies on all six days, the coupling-off twin
and the degraded edge are TOST-equivalent to zero, readout unfolding raises the value but never carries the
claim, and the warm-start arm certifies below the cold value.

\begin{table}[t]
\centering
\caption{\textbf{Arm-by-arm ablation on \texttt{b08}.} Day-level grand means over six calibrations. The
negativity, the $J=0$ twin and the degraded $6$-$8$ edge carry PSD+BCa intervals; the concurrence,
log-negativity and CHSH-operator maximum carry day-level intervals as robustness cross-checks. TOST
equivalence is at margin $\delta=0.02$.}
\label{tab:ablation}
\begin{tabular}{lcc}
\toprule
arm / quantity & value & $95\%$ interval (type) \\
\midrule
good $4$-$5$, $p{=}1$, raw (certificate) & $0.077$ & $[0.065,0.091]$ (BCa) \\
good $4$-$5$, $p{=}1$, readout-mitigated (IBU) & $0.152$ & $[0.138,0.170]$ (BCa) \\
warm-start good $4$-$5$, $p{=}1$ & $0.057$ & $[0.048,0.066]$ (hier.\ BCa) \\
coupling-off $J{=}0$ twin & $0.000$ & TOST${=}0$ (upper $0.0001$) \\
degraded $6$-$8$, $p{=}1$ & $0.001$ & $[0.000,0.007]$ (BCa); TOST${=}0$ (upper $0.0046$) \\
good $4$-$5$, $p{=}3$ (depth collapse) & $0.000$ & TOST${=}0$ (upper $0.000$) \\
concurrence, good $4$-$5$, $p{=}1$ & $0.165$ & $[0.131,0.200]$ (day-level) \\
log-negativity, good $4$-$5$, $p{=}1$ & $0.207$ & $[0.164,0.250]$ (day-level) \\
CHSH-operator max, good $4$-$5$, $p{=}1$ & $1.37$ & $[1.31,1.44]$ (day-level) \\
\bottomrule
\end{tabular}
\end{table}

\paragraph{Randomized-measurement power and stopping rules.} The witness threshold is the $95$th
percentile of $\hat W$ under a Werner PPT-boundary null; on Red the measured $W=0.0105$ lies $7.2$ null
standard deviations above that threshold ($t_{95}=0.0027$) and the empirical false-positive rate of the
CI-excludes-zero rule is $1.5\%$, with the simulated null mean $W$ within $10^{-6}$ of zero. The honest
null, however, is the near-pure-separable family, whose finite-sample moment variance is large: at the
Werner-equivalent magnitude realised on Red ($W\approx0.016$), pushing the $95$th percentile of that null
below the measured signal requires $N_U\approx12{,}672$ random-unitary settings, against the $N_U=800$
actually run, so no feasible threshold both controls the false-positive rate and retains power. The same
estimator is fully powered at the ideal magnitude ($W\approx0.245$, detection power $\approx1$). This is
the basis for reporting the on-Red witness as consistent-with-entanglement and elevating the negativity to
the primary certificate. The certificate's own stopping rule was pre-registered as: add calibration days
until the day-level grand-mean CI half-width falls to $0.02$ or below; the six-day window met it at a
half-width of $0.017$.

\paragraph{Per-day calibration and readout characterisation.} Each day we measured the per-qubit readout
confusion in both directions and a Bell $Z$-correlator on each used edge, hashed and bound to that day's
SLURM job identifiers. On day one the readout confusion $(p(1|0),p(0|1))$ was $q_4\,(0.011,0.078)$,
$q_5\,(0.060,0.049)$, $q_6\,(0.008,0.052)$, $q_7\,(0.011,0.071)$, $q_8\,(0.007,0.032)$, and the Bell
$Z$-correlator $P(\text{equal})$ was $0.894$ on the good edge and $0.825$ on the degraded edge. The
six-day autocalibration CZ fidelities averaged $94.7\%$ (good $4$-$5$) and $74.7\%$ (degraded $6$-$8$),
with $94.9\%/74.0\%$ on the randomized-measurement day; single-shot readout was $\approx96\%$. The per-day
negativities behind the certificate are in Table~\ref{tab:perday}; the per-day readout confusion for days
two to six and the mitigated-versus-raw confusion matrices are in the deposit.

\paragraph{Withdrawn and deferred runs.} We record what was not used, so the scope is auditable. The prior
five-qubit attempt is withdrawn in full: its cost layer compiled to the diagonal $\mathrm{CZ}\,R_Z\,\mathrm{CZ}=R_Z$
product with no ZZ coupling, so every absolute number from it is retracted and motivates the per-circuit
statevector guardrail. Deferred in this reduced pilot: the ZNE-folding mitigation rung (not executed); the
scrambled-encoding hardware null for the optimum-mass test (so PP2 is only partially tested); the
line-three mechanism observables (connected correlators, per-edge mutual information) and the global-purity
control (not processed); and the depth-one-and-two ensemble with its out-of-sample noise model. One
auxiliary cross-architecture witness computed with ion-trap smoke-test tooling ($W=0.4757$, meaningless
confidence interval) was discarded as not fit for inference and plays no role in any claim.

\paragraph{Causal experiment: per-run results.} The dose-response is fitted on the \emph{manipulated} CZ
pulse amplitude factor, because the measured average gate fidelity $F_\mathrm{avg}$ is non-monotone in the
drive (its Ramsey conditional phase wraps) and does not track the certified negativity. Table~\ref{tab:causal}
gives the six per-run amplitude-axis slopes with rung-level lack-of-fit-honest intervals, their coefficient
of determination $R^2$ and reduced $\chi^2$, the CZ-off twin negativity, and the coupled-rung fringe
visibility band. Table~\ref{tab:causaldiag} documents, for the same six runs, that the $F_\mathrm{avg}$ axis
fails (weak, non-monotone correlation and a reduced $\chi^2\approx100$--$137$ that no straight line
absorbs) while the physically equivalent conditional-phase axis $\lvert\zeta\rvert$ succeeds, and records the
twin's mid-axis $F_\mathrm{avg}$. Table~\ref{tab:causalpool} gives the per-edge pooled slopes and the
edge-level random-effects summary. The re-analysis is computed from each record's raw per-rung amplitude and
negativity arrays; its self-check reproduces every sealed \texttt{regression.slope} (the $F_\mathrm{avg}$
fit) to $0.0\times10^0$ and matches every sealed \texttt{regression.boot\_sd} with its analytic shot-level
slope error to $<5\times10^{-4}$, confirming the pipeline before the axis is corrected. The $\lvert\zeta\rvert$
column is computed from each record's per-rung \texttt{fidelity.zeta} (stored modulo $2\pi$) by physical
unwrapping along the amplitude ladder, and is emitted by the same script so it is reproducible.

\begin{table}[t]
\centering
\caption{\textbf{Causal dose-response on the manipulated drive, per run.} Slope of certified negativity on
the manipulated CZ pulse amplitude factor (certified negativity per unit fractional amplitude), with a
rung-level lack-of-fit-honest $95\%$ interval (shot-level slope error inflated by
$\sqrt{\smash[b]{\chi^2_\nu}}$, Student-$t$ on $\nu=6$), the fit $R^2$ and reduced $\chi^2$, the CZ-off
$J{=}0$ twin negativity (PSD), and the coupled-rung mean fringe visibility band, for three couplers across
two calibration sessions. Every run has a positive slope with a $95\%$ interval excluding zero
(one-sided $p<10^{-3}$).}
\label{tab:causal}
\begin{tabular}{llccccc}
\toprule
coupler & session & slope [rung-honest $95\%$] & $R^2$ & red.\ $\chi^2$ & twin neg. & visibility \\
\midrule
$4$-$5$ & s1 & $8.14\ [6.55,9.73]$   & $0.96$ & $4.4$  & $0.000$ & $0.82$--$0.86$ \\
$4$-$5$ & s2 & $8.70\ [6.61,10.80]$  & $0.95$ & $7.8$  & $0.000$ & $0.82$--$0.87$ \\
$5$-$7$ & s1 & $8.14\ [5.72,10.56]$  & $0.92$ & $10.4$ & $0.001$ & $0.83$--$0.88$ \\
$5$-$7$ & s2 & $9.49\ [6.17,12.80]$  & $0.89$ & $15.5$ & $0.000$ & $0.83$--$0.89$ \\
$6$-$8$ & s1 & $8.97\ [5.18,12.75]$  & $0.85$ & $21.0$ & $0.007$ & $0.77$--$0.81$ \\
$6$-$8$ & s2 & $8.54\ [4.72,12.35]$  & $0.83$ & $21.6$ & $0.005$ & $0.76$--$0.80$ \\
\bottomrule
\end{tabular}
\end{table}

\begin{table}[t]
\centering
\caption{\textbf{Why $F_\mathrm{avg}$ is a diagnostic, not the regressor.} For each run: the discarded
fidelity axis (Pearson $r$ of certified negativity with $F_\mathrm{avg}$, and the weighted $R^2$ and reduced
$\chi^2$ of the straight-line fit, and the CZ-off twin's $F_\mathrm{avg}$, which sits mid-axis rather than at
zero), and the physically equivalent conditional-phase axis $\lvert\zeta\rvert$ (slope in certified
negativity per radian, weighted $R^2$). $R^2$ is the weighted coefficient of determination throughout, the
same definition used in Table~\ref{tab:causal}, for parity; $\lvert\zeta\rvert$ is the per-rung conditional
phase physically unwrapped along the amplitude ladder (the stored \texttt{fidelity.zeta} wraps modulo
$2\pi$). With rung-honest errors no $F_\mathrm{avg}$ slope excludes zero; every $\lvert\zeta\rvert$ slope
does. $F_\mathrm{avg}$ was swept over $\approx0.20$--$1.00$ on every run.}
\label{tab:causaldiag}
\begin{tabular}{llcccc c}
\toprule
 & & \multicolumn{4}{c}{$F_\mathrm{avg}$ axis (discarded)} & $\lvert\zeta\rvert$ axis \\
\cmidrule(lr){3-6}\cmidrule(lr){7-7}
coupler & session & Pearson $r$ & $R^2$ & red.\ $\chi^2$ & twin $F_\mathrm{avg}$ & slope [$R^2$] \\
\midrule
$4$-$5$ & s1 & $0.29$ & $0.14$ & $104$ & $0.40$ & $0.092\ [0.90]$ \\
$4$-$5$ & s2 & $0.32$ & $0.14$ & $122$ & $0.40$ & $0.097\ [0.86]$ \\
$5$-$7$ & s1 & $0.36$ & $0.18$ & $105$ & $0.40$ & $0.090\ [0.86]$ \\
$5$-$7$ & s2 & $0.20$ & $0.06$ & $134$ & $0.40$ & $0.086\ [0.77]$ \\
$6$-$8$ & s1 & $0.15$ & $0.01$ & $137$ & $0.37$ & $0.084\ [0.79]$ \\
$6$-$8$ & s2 & $0.12$ & $0.01$ & $128$ & $0.36$ & $0.084\ [0.84]$ \\
\bottomrule
\end{tabular}
\end{table}

\begin{table}[t]
\centering
\caption{\textbf{Causal dose-response on the manipulated drive, pooled.} Per-coupler slopes pool the two
sessions by inverse-variance weighting of the two rung-honest session slopes (units: certified negativity
per unit fractional CZ amplitude); the one-sided per-coupler tests have $p<10^{-3}$ and survive Holm and
Benjamini--Hochberg correction across the three couplers. On the manipulated axis the couplers are
homogeneous ($Q=0.15$, df $2$, $I^2=0\%$, $\tau^2=0$; at $k=3$ couplers these heterogeneity estimates are
near-uninformative), so the DerSimonian--Laird and fixed-effect pools coincide. The conservative
small-sample Knapp--Hartung--Sidik--Jonkman interval also excludes zero once its dispersion factor is floored
to one ($[6.73,10.21]$); we floor it because the raw factor $q=0.076<1$ signals underdispersion at $k=3$, so
the unfloored KHSJ interval ($[7.99,8.95]$) is anti-conservative and is shown only for reference. This is in
contrast to the discarded $F_\mathrm{avg}$ regression, whose apparent heterogeneity ($Q=56.2$, $I^2=96.4\%$)
was an artefact of the non-monotone axis.}
\label{tab:causalpool}
\begin{tabular}{lcc}
\toprule
coupler / pool & slope [$95\%$ CI] & one-sided $p$ (Holm, BH) \\
\midrule
$4$-$5$ & $8.35\ [7.33,9.36]$ & $<10^{-3}$ (survive, survive) \\
$5$-$7$ & $8.61\ [7.04,10.17]$ & $<10^{-3}$ (survive, survive) \\
$6$-$8$ & $8.75\ [6.60,10.91]$ & $<10^{-3}$ (survive, survive) \\
\midrule
random-effects (DL, normal) & $8.47\ [7.68,9.26]$ & excludes $0$ \\
random-effects (KHSJ, $k{=}3$, $q$ floored to $1$) & $8.47\ [6.73,10.21]$ & excludes $0$ \\
\quad(KHSJ unfloored, $q=0.076$, anti-conservative) & $8.47\ [7.99,8.95]$ & (reference only) \\
fixed-effect (reference) & $8.47\ [7.68,9.26]$ & excludes $0$ \\
\bottomrule
\end{tabular}
\end{table}

\paragraph{Causal experiment: pre-registration and external timestamps.} Each of the six per-session
plans was sealed as a \texttt{PREREGISTRATION.json} and independently timestamped by an RFC-3161 Time-Stamp
Authority (freetsa.org) before that day's data existed. The tokens were verified locally
(\texttt{openssl ts -reply}), all with status \emph{Granted}: the three session-2 plans at
2026-09-02 01:54:01--03\,Z and the three session-1 plans at 2026-09-01 08:23:31--54\,Z, each token's
message imprint equal to the SHA-256 of its plan file (edge-$4$-$5$ session-2 imprint
\texttt{e778820f\ldots} $=$ that file's SHA-256; the six SHA-256 seals are listed in the Data-availability
section). Data collection for each session began after its seal (session-1 fine sweep at
$\approx$09:36\,Z on 2026-09-01, session-2 at $\approx$05:42\,Z on 2026-09-02), so the design preceded the
data. The master runcards were read-only, their SHA-256 re-verified identical before and after every rung
(session-1 \texttt{e8db3ca5\ldots}, session-2 \texttt{d597a096\ldots}).

\bibliographystyle{quantum}
\bibliography{refs}

\end{document}